\documentclass[reprint, notitlepage, nofootinbib, twocolumn]{revtex4-1}

\usepackage{t1enc}
\usepackage[latin2]{inputenc}
\usepackage{color}

\usepackage{savesym}
\usepackage{txfonts}
\usepackage{tipa}
\savesymbol{iint}
\savesymbol{iiint}
\savesymbol{iiiint}
\savesymbol{idotsint}
\usepackage{amsmath,amsfonts,amssymb}
\usepackage{mathscinet, mathtools}
\savesymbol{degree}
\restoresymbol{TXF}{iint}
\restoresymbol{TXF}{iiint}
\restoresymbol{TXF}{iiiint}
\restoresymbol{TXF}{idotsint}
\restoresymbol{STD}{degree}
\usepackage{MnSymbol}
\usepackage{nicefrac}
\usepackage{hyperref}
\usepackage{bbm}

\usepackage{multirow}
\usepackage{hyperref}
\usepackage{array}

\usepackage{graphics}
\usepackage{float, subfig}
\usepackage{graphicx}
\usepackage{epsfig, caption}

\newcommand{\ve}[1]{\mathbf{#1}}

\definecolor{!R}{rgb}{1,0,0}
\definecolor{!G}{rgb}{0,1,0}
\definecolor{!B}{rgb}{0,0,1}
\definecolor{!W}{rgb}{1,1,1}
\definecolor{orange}{RGB}{255,128,0}
\newcommand{\rem}[1]{{#1}}
\newcommand{\remB}[1]{{#1}}
\newcommand{\remC}[1]{{#1}}

\newcommand{\avr}[1]{\langle #1\rangle}
\newcommand{\intlim}[3]{\int\limits_{#2}^{#3}\!\!\mathrm{d}#1}
\newcommand{\intlimB}[3]{\int\limits_{#2}^{#3}\!\!\text{\textcrd}#1}

\usepackage{slashed}
\usepackage{empheq} 
\usepackage{calrsfs}
\DeclareMathAlphabet{\mathpzc}{OT1}{pzc}{m}{it}

\newcommand{\wt}[1]{\widetilde{#1}}

\newcommand{\specLB}{{\color{!W}{\intlim{}{0}{0}}}}

\newcommand{\nn}{\nonumber}

\newcommand{\actSign}{\REM{-}}
\newcommand{\ChIm}{\REM{\mu_{5}}}
\newcommand{\FTChIm}{\REM{\wt{\mu}_{5}}}
\newcommand{\OLChIm}{\REM{\overline{\mu}_{5}}}

\newcommand{\citeRev}{\cite{anomTransportRev,LandsteinerRev,FukushimaLocalCP,FukushimaRev2,ReviewBook,JLrev}}
\newcommand{\citeHICmes}{\cite{polSTAR1,polSTAR2,polSTAR3,HICmes1}}
\newcommand{\citeCMmes}{\cite{CondmatMes1,CondmatMes2,CondmatMes3,CondmatMes4,CondmatMes5,CondmatMes6}}
\newcommand{\citeCKT}{\cite{Shovkovy6,Shovkovy7,Shovkovy8,chiralPlasma1,chiralPlasma2,nonLinRespKinTh1,nonLinRespKinTh2}}
\newcommand{\citeCKTonlyRHIC}{\cite{chiralPlasma1,chiralPlasma2,nonLinRespKinTh1,nonLinRespKinTh2,WignerFunc1,WignerFunc2,WignerFunc3,nsCMEKharzeev2}}
\newcommand{\citeWigner}{\cite{WignerFunc1,WignerFunc2,WignerFunc3}}
\newcommand{\citeQFT}{\cite{nsCMEKharzeev,nsCMEKharzeev2,AnomCondImagTime,HLR2011,secondOrderTransp,CMErelaxFLTfermions,WHR2017}}
\newcommand{\citeHolo}{\cite{LublinskyHolo1,LublinskyHolo2,LandsteinerHolo1,LandsteinerHolo2,HoloQuenches,YeeHolo,Rubakov}}
\newcommand{\citeNoEqCME}{\cite{HoloQuenches,Rubakov,FukushimaNonstatCME,YamamotoBlochT,Zubkov1,Zubkov2}}
\newcommand{\citeNoCorr}{\cite{CMEradcorrQED,NonrenormCMEpQCD,CMEnocorrQED,Zubkov1,Zubkov2}}

\newcommand{\affHuashi}{\affiliation{Institute of Particle Physics and Key Laboratory of Quark and Lepton Physics (MOE), \\ Central China Normal University, Wuhan 430079, China}}
\newcommand{\affRockefeller}{\affiliation{Physics Department, The Rockefeller University, 1230 York Avenue, New York, New York 10021-6399, USA}}
\newcommand{\affIndiana}{\affiliation{Physics Department and Center for Exploration of Energy and Matter, Indiana University, 2401 N Milo B. Sampson Lane, Bloomington, Indiana 47408, USA}}

\newcommand{\REM}[1]{#1}
\newcommand{\tobs}{\tau_o}

\begin{document} 
\title{Chiral magnetic response to arbitrary axial imbalance}
\author{Mikl\'os Horv\'ath}
\email[e-mail: ]{miklos.horvath@mail.ccnu.edu.cn}
\affHuashi
\author{Defu Hou}
\email[e-mail: ]{houdf@mail.ccnu.edu.cn}
\affHuashi
\author{Jinfeng Liao}
\email[e-mail: ]{liaoji@indiana.edu}
\affIndiana
\author{Hai-cang Ren}
\email[e-mail: ]{ren@mail.rockefeller.edu}
\affRockefeller \affHuashi

\begin{abstract}
The response of chiral fermions to time and space dependent axial imbalance \& constant magnetic field is analyzed. The axial-vector--vector--vector {{(AVV)}} three-point function is studied using a real-time approach at finite temperature in the {weak external field} approximation. The chiral magnetic conductivity is given analytically for noninteracting fermions. It is pointed out that local charge conservation plays an important role when the axial imbalance is inhomogeneous. Proper regularization is needed which makes the constant axial imbalance limit delicate: for static {{but spatially oscillating}} chiral charge the current of the chiral magnetic effect (CME) vanishes. In the homogeneous (but possible time-dependent) limit of the axial imbalance the CME current is determined solely by the chiral anomaly. As a phenomenological consequence, the observability of the charge asymmetry caused by the CME turns out to be a matter of interplay between various  scales of the system. Possible plasma instabilities resulting from the gradient corrections to the CME current are also pointed out.  
\end{abstract}

\newpage
\maketitle

\section{Introduction}
Anomaly induced transport phenomena in systems with chiral fermions have attracted wide interests ranging from high energy physics to condensed matter physics. Among them is the chiral magnetic effect (CME) which relates the chiral chemical potential $\mu_5$ and the external magnetic field ${\bf B}$ to the anomaly induced electric current density ${\bf J}$ by the simple formula \cite{KharzeevOriginalCME,KharzeevOriginalCMElonger}
\begin{align}
\ve{J}=&\frac{e^2}{2\pi^2}\mu_5\ve{B}.
\label{CME}
\end{align}
The predictions of CME include the electric charge asymmetries in the final stage of the relativistic heavy ion collisions (RHIC) \citeHICmes and the negative magnetoresistance in some Weyl and Dirac semimetals \citeCMmes. In the former case, a strong magnetic field is generated during an off-central collision and the chirality imbalance is induced by the transition among different topological sectors. Therefore, the CME is an important probe of the topological structure of QCD. While there are experimental evidences of CME in the context of condensed matter physics, the situation in RHIC is far more complicated. It remains to exclude the noisy backgrounds in order to nail down the real CME signals. 

For the past decade since the concept of CME was proposed there have been a vast amount of theoretical works done on the subject. For thorough reviews, see Refs. \citeRev and the references therein. \REM{For a recent review on the status of CME in RHIC see the relevant parts of Ref. \cite{JLrev}.}
Considering that the CME supposed to have a macroscopic imprint, hydrodynamic descriptions including the effect of the anomaly have been developed in order to simulate the modified dynamics of the medium \cite{SonSorowka,AVFD1,AVFD2}. The underlying assumption when applied to RHIC is that a net macroscopic chiral charge is generated in the initial stage of collisions and its characteristic time of variation is much longer than the relaxation time required to establish a local thermal equilibrium, so the formula (\ref{CME}) can be applied. Hydrodynamic modeling of anomalous transport in condensed matter systems has been actively investigated as well \cite{Shovkovy1,Shovkovy2,Shovkovy3,Shovkovy4}.

There are several other ways to approach the transport phenomena starting from the microscopic level. Investigations have been conducted ranging from kinetic theory (Boltzmann equations \citeCKT  or Wigner functions \citeWigner) to field theoretic approaches (Kubo formulas) \citeQFT, even through holographic models \citeHolo for an insight of strongly coupled systems. {{In the equilibrium case,}} all of them lead to the same answer as given in \mbox{Eq. (\ref{CME})}.

{\remC{
The chiral magnetic response in the nonequilibrium case, in particular for a space-time-dependent chiral chemical potential, turns out to be both subtle and important, for a number of reasons. First of all, in the context of heavy ion collisions, the initial axial charge is generally expected to be inhomogeneous across the fireball and furthermore necessarily evolves in time due to random gluonic topological transitions during the fireball evolution. The spatial variation length scale and the time evolution scale are not necessarily very large as compared with the thermal scales of the medium. It is therefore crucial to understand the impact of such nonequilibrium case for application to phenomenology (e.g. for the charge asymmetry signal of CME in these collisions). Secondly,  as was pointed out in \mbox{Ref. \cite{WHR2017}} using proper UV regularization, connecting chiral magnetic response under the space-time dependent chiral chemical potential with that under static and homogeneous chiral chemical potential could be tricky. Depending on the order of taking static limit first or taking homogeneous limit first, the results are completely different. What that implies for a realistic system like the quark-gluon plasma in heavy ion collisions, has remained unclear so far. These are the pressing issues that we plan to explore in the present work.}}

\rem{
{\remC{In order to do this, we employ a relatively simple but clean theoretical setup, by considering  the weak external field approximation (WFA) of a fermionic system coupled to electromagnetic (EM) fields as well as under the presence of space-time dependent chiral chemical potential.}} We shall derive an explicit formula for the chiral magnetic current with an arbitrary spacetime dependent $\mu_5$ in a constant magnetic field. 
Based on such results, we shall then explore its impact on the charge asymmetry transport across a plane that is perpendicular to the magnetic field, which mimics the situation of chiral magnetic transport across the reaction plane in RHIC.
}

The rest of this paper is organized as follows. The Schwinger-Keldysh formulation for the amplitude of the triangle diagram, retarded with respect to the chiral chemical potential and external magnetic field are laid out in the next Section \ref{secVertex}. The explicit form of the response function under a constant magnetic field is presented in Section \ref{secConstB} and its contribution to the electric charge asymmetry is discussed in Section \ref{secChargeAsymm}. Section \ref{secConclusions} concludes the paper.

\section{Weak external field approximation}\label{secVertex}
In this section we introduce the weak external field approximation (WFA) of a fermionic system coupled to electromagnetic fields as well as in the presence of axial-vector potential. Such a model might capture the electromagnetic transport in a quark-gluon plasma of which the gluon sector can have nontrivial topological features, locally violating $CP-$invariance. This local $CP-$violation is described through the axial-vector potential $A_5$ under our assumption. On the fundamental level of QCD there are no axial gauge fields, however, focusing only on the effective description of the EM sector, there are two contributions to the chiral charge nonconservation. The usual EM one proportional to $\ve{E}\cdot\ve{B}$, and the one from the gluonic sector: $\propto \ve{E}_a\cdot\ve{B}^a$, with $\ve{E}_a$ and $\ve{B}_a$ being the components of chromoelectric and -magnetic fields, respectively. It is the latter contribution that is reflected in the hypothetical axial-vector potential.

Solid state systems might happen to be affected by similar circumstances, such as electronic systems in the bulk of Weyl semimetals (WSM). Their low energy behavior is described by the Dirac like equation. Since the spatial inversion and time reversal symmetries can be violated in such systems, depending on the details of the material, one can introduce an axial-vector potential. It was explicitly shown in Ref. \cite{InducedAxialFields} that the elastic deformations of a WSM sample can be modeled in terms of an axial-vector potential as well. See for example Ref. \cite{LandsteinerAxialTr}, Sec. 7 of Ref. \cite{LandsteinerRev} or Secs. 2, 5 of Ref. \cite{Shovkovy1} for more details. 

The EM sector of the underlying quantum field theory can be described by the following Lagrangian:

\begin{align}
\mathcal{L} =& -\frac{1}{4}F_{\mu\nu}F^{\mu\nu} + \overline{\psi}\gamma^\mu\left(i\partial_\mu -eA_\mu -\gamma^5A_{5,\mu}\right)\psi + \text{ UV reg. }\nonumber \\
 =& \mathcal{L}_\text{QED} -\overline{\psi}\gamma^\mu\gamma^5\psi\cdot A_{5,\mu}, \label{EMLagrange}
\end{align} 
with the Dirac field $\psi$ and the electromagnetic vector potential $A_\mu$. It is worthwhile to mention that for $A_5=(-\mu_5,\ve{0})$ the coupling to the axial-vector field effectively behaves as an axial chemical potential, i.e., $A_{5,0}=-\mu_5$. These are, however, two fundamentally different concepts, since there is no need to impose the constraints of thermal equilibrium in order to talk about an axial-vector potential as a proxy to axial charge imbalance. 

\begin{figure}
\includegraphics[width=0.55\linewidth]{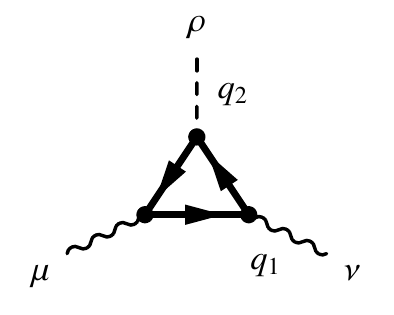}
\caption{One-loop triangle diagram with two vector and an axial-vector vertex, indexed by $\mu$, $\nu$, and $\rho$, respectively.}\label{fig:triangleDiagram}
\end{figure}

\rem{
In order to investigate the dynamics of anomalous chiral transport we expand the currents of interest up to the first nontrivial order with respect to the external fields: to capture anomaly induced transport, this means $AA$, $AA_5$ and $A_5A_5$ need to be included. Since $A$ and $A_5$ are treated as classical background fields, this procedure is justified as a \textit{weak external field expansion }
and it can be realized as follows (repeated indexes are summed up):
}
\begin{align}
\avr{J^\mu} =& \textstyle \left.\avr{J^\mu J^\nu}\right|_{A,A_5=0}A_\nu + \left.\avr{J^\mu J^\nu J^\rho_5}\right|_{A,A_5=0}A_\nu A_{5,\rho} +\dots, \label{Jlinres}\\
\avr{J_5^\mu} =& \textstyle \left.\avr{J^\mu_5 J^\nu_5}\right|_{A,A_5=0}A_{5,\nu} +\frac{1}{2}\left.\avr{J_5^\mu J^\nu J^\rho}\right|_{A,A_5=0}A_\nu A_\rho + \nonumber \\
& +\textstyle \frac{1}{2}\left.\avr{J_5^\mu J_5^\nu J_5^\rho}\right|_{A,A_5=0}A_{5,\nu} A_{5,\rho} +\dots, \label{J5linres}
\end{align}
with the (axial) vector potential $A^\mu_{(5)}$. \rem{The ensemble average $\avr{\dots}$ of the (axial)vector current $\avr{J^\mu_{(5)}}=\avr{\overline{\psi} \gamma^\mu(\gamma^5)\psi}$ is to be taken over a thermal ensemble of quantum states, characterized by zero external fields.} 
Both \mbox{Eqs. (\ref{Jlinres})} and (\ref{J5linres}) are to be understood as convolutions in the space-time arguments for inhomogeneous and time dependent sources. In the special case of zero electric field and zero axial magnetic field, the electric current $\ve{J}$ is solely controlled by the response function $\avr{J^iJ^jJ_5^0}$. The functions $\avr{JJJ_5}$ contains two vector and one axial-vector current operators and will be referred to as axial-vector--vector--vector (AVV) response from here on. From the field theoretic perspective, this function corresponds to the fully dressed triangle diagram (see Fig. \ref{fig:triangleDiagram}) whose longitudinal component with respect to the $A$-vertex is --- in the constant field limit --- dictated by the chiral anomaly. As we will see in the followings, specific orders of limits are tied to the logitudinal component, but this is not the case generally, as it was already pointed out in Ref. \cite{HLR2011}.

The contribution of the second term in \mbox{Eq. (\ref{Jlinres})} of the electric current is given by the following expression:
\begin{align}
\avr{J^i}(x) =& \intlim{^4y}{}{}\intlim{^4z}{}{}A_j(y)A_{5,0}(z)\Gamma^{0ij}_{AVV}(x-y,x-z) =\nonumber \\
=& \intlimB{^4q_1}{}{}\intlimB{^4q_2}{}{}\wt{A}_j(q_1)\wt{A}_{5,0}(q_2)\wt{\Gamma}^{0ij}_{AVV}(q_1,q_2)e^{ix\cdot(q_1+q_2)}. \label{responseCurrent}
\end{align}
We again point out that for any explicit calculation one needs to regularize the theory in order to keep the electromagnetic $U(1)$ Ward--Takahasi-identity intact. Because of the presence of $\gamma^5$, we use the method of Pauli and Villars, 
i.e. coupling to the system an auxiliary field with asymptotically large mass $M$, which although obeys Fermi-Dirac statistics, contributes with an opposite sign to the loop integrals. So in addition to the usual AVV triangle we have to subtract the one with the heavy fermions. 

In order to investigate the response current out of equilibrium we need to formulate the AVV triangle in terms of the real-time correlations of the underlying field theory. Using the Schwinger-Keldysh (SK) formalism for this purpose, the Fourier transformed AVV vertex $\wt{\Gamma}^{\rho\mu\nu}_{AVV}$, responsible for the retarded current response, then reads as follows \cite{WHR2017}:

\begin{align}
\wt{\Gamma}^{\rho\mu\nu}_{AVV}(q_1,&q_2) = \nonumber \\
 -\frac{ie^2}{2}\int_p \text{tr}& \left\{ \gamma^\mu G^C(p+q_2)\gamma^\rho\gamma^5 G^A(p) \gamma^\nu G^A(p-q_1) +\right. \label{GAVVline1} \\
& +\gamma^\mu G^R(p+q_2)\gamma^\rho\gamma^5 G^C(p) \gamma^\nu G^A(p-q_1) +\label{GAVVline2}\\
& +\gamma^\mu G^R(p+q_2)\gamma^\rho\gamma^5 G^R(p) \gamma^\nu G^C(p-q_1) +\label{GAVVline3}\\
& +\gamma^\mu G^C(p+q_1)\gamma^\nu G^A(p) \gamma^\rho\gamma^5 G^A(p-q_2) + \label{GAVVline4}\\
& +\gamma^\mu G^R(p+q_1)\gamma^\nu G^C(p) \gamma^\rho\gamma^5 G^A(p-q_2) +\label{GAVVline5}\\
& +\left. \gamma^\mu G^R(p+q_1)\gamma^\nu G^R(p) \gamma^\rho\gamma^5 G^C(p-q_2) \right\}  \label{GAVVline6}\\
& -\left\{\text{same terms with mass }M\gg\text{all other} \right. \nonumber \\
& \left.\text{ scales of the system}\right\},\label{GAVVline7}
\end{align}
where $G^{R,A,C}$ are the retarded, advanced, and correlation components of the fermionic propagator in the SK formalism, respectively. \REM{The above vertex function is retarded with respect to its index $\mu$, so the response current follows the perturbations both in $A$ and $A_5$.} For a detailed introduction into the formalism see for example Ref. \cite{SchwKelFormalism1}. All the propagators are to be considered at zero external fields in thermal equilibrium. All components are linked to the fermionic spectral density $\rho=iG^R-iG^A$, $iG^C(p)=\rho(p)\cdot(1-2\wt{n}(p_0))$, where we suppressed the temperature dependence of the Fermi-Dirac distribution $\wt{n}(p_0)=n_{FD}(p_0/T)$. For the subsequent calculations, we also introduce $\wt{\Gamma}^{\rho\mu\nu}_{AVV} =:\mathfrak{g}^{\rho\mu\nu}_m -\mathfrak{g}^{\rho\mu\nu}_\text{PV},$ where the Pauli-Villars-term (PV) is $\mathfrak{g}^{\rho\mu\nu}_\text{PV}(q_1,q_2)\equiv\mathfrak{g}^{\rho\mu\nu}_{m=M}(q_1,q_2)$, $M$ being much larger than any other scales $q_1$, $q_2$, $m$ or $T$. So practically $M$ is sent to infinity while other external parameters are kept finite. 
The AVV vertex satisfies the following equations due to the the Ward--Takahasi-identities:

\begin{align}
(q_{1}+q_2)_\mu\wt{\Gamma}_{AVV}^{\rho\mu\nu} =& q_{1,\nu}\wt{\Gamma}_{AVV}^{\rho\mu\nu} =0, \label{gradJ}\\
q_{2,\rho}\wt{\Gamma}_{AVV}^{\rho\mu\nu} =& i\epsilon^{\mu\nu\alpha\beta}q_{1,\alpha}q_{2,\beta}\cdot\frac{e^2}{2\pi^2}. \label{gradJ5}
\end{align}
The first two equations in \mbox{Eq. (\ref{gradJ})} implies $\partial\cdot J=0$ at the level of the approximation scheme, whilst \mbox{Eq. (\ref{gradJ5})} is the anomalous nonconservation of the axial-vector current $J_5$. The above properties of the AVV vertex are the consequence of the following identities:

\begin{align}
G^{R,A}(p+q)\slashed{q}G^{R,A}(p) =& -G^{R,A}(p+q) + G^{R,A}(p), \label{WardG1}\\
G^C(p+q)\slashed{q}G^{R,A}(p) =& -G^C(p+q), \label{WardG2}\\
G^{R,A}(p+q)\slashed{q}G^{C}(p) =& G^C(p). \label{WardG3}
\end{align}
Plugging \mbox{Eq. (\ref{gradJ5})} into \mbox{Eq. (\ref{J5linres})} one obtains the well-known anomalous Ward-identity in the chiral limit:
\begin{align}
\avr{\partial\cdot J_5} =& \frac{1}{2}\intlimB{^4q_1}{}{}\intlimB{^4q_2}{}{}\wt{A}_\mu(q_1)\wt{A}_\nu(q_2)\times \nn \\
& i(q_1+q_2)_\rho\wt{\Gamma}_{AVV}^{\rho\mu\nu}(q_2,-q_1-q_2)e^{ix\cdot(q_1+q_2)} =\nn\\
=&\frac{e^2}{2\pi^2}\ve{E}\cdot\ve{B}.
\end{align}
Setting $\ve{q}_2=0$, $\wt{\Gamma}_{AVV}^{0\mu\nu}(q_1,q_2)$ itself becomes completely determined by the UV sector of the theory, i.e. ruled by the anomaly. Formally this behavior is caused by cancellation between certain terms in the vertex function which are equal upon a shifting of the loop momentum. The details can be found in Appendix \ref{app:ward}. For the electric current this means the chiral magnetic effect prevails even for \textit{time-dependent but homogeneous $\ChIm$} and arbitrary $\ve{B}$:
\begin{align}
\ve{J}(t,\ve{r})=& \frac{e^2}{2\pi^2}A_{5,0}(t)\ve{B}(t,\ve{r}) =-\frac{e^2}{2\pi^2}\mu_5(t)\ve{B}(t,\ve{r}).\label{CME1}
\end{align}
\rem{
This anomaly driven chiral magnetic current in \mbox{Eq. (\ref{CME1})} differs from that in \mbox{Eq. (\ref{CME})} by a sign
at a constant $\mu_5$ and applies to a magnetic field with arbitrary spacetime dependence. 
\remB{We point out that this sign difference stems from \textit{the inclusion or absence of the PV term besides the contribution of the triangle diagram}. As the inclusion results zero CME current in the static $\mu_5$ limit -- and to \mbox{Eq. (\ref{CME1})} in the homogeneity limit --, while the absence of it leads to \mbox{Eq. (\ref{CME})}. }\remC{It is also important to point out that the role of the PV term is UV-regularization: it keeps the electric (vector)charge conservation $\partial\cdot J=0$ intact. Otherwise, there is charge generation at the boundary of the system, proportional to gradients of the axial imbalance. We note here that in order to ensure local charge conversation it is equivalently possible to add a so-called Bardeen-Zumino counterterm or Chern-Simons (CS) term to the effective action. The electric current is then going to have an additional contribution -- the Chern-Simons current --, which cancels the traditional CME current of \mbox{Eq. (\ref{CME})}.

The absence of PV term in several kinetic theory works from the high-energy side \mbox{\citeCKTonlyRHIC} and the early field theory analysis like in \mbox{Refs. \cite{KharzeevOriginalCME, KharzeevOriginalCMElonger}} lead to a delayed response current for time-dependent external fields. When included, however, the CS current causes the response to have an instantaneous contribution, at least in the weak coupling approximation. 
The inclusion or exclusion of the CS term reflects the different definitions of the electric current as an operator: the \textit{consistent anomaly} ensures the vector charge to be conserved, while the \textit{covariant anomaly} defines $J$ to transform covariantly under any gauge transformation \cite{LandsteinerRev}. It is important to point out, that in the context of condensed matter physics, the derivation of the low-energy theory from the lattice regularized one will lead essentially to a CS term, as pointed out in Ref. \cite{Shovkovy5}. }

As is shown in Appendix B, the sign of \mbox{Eq. (\ref{CME1})} matches the sign of the chiral magnetic current of the Maxwell-Chern-Simons electrodynamics: when the $\wt{F} F$ term of the latter is converted to the 
corresponding term \mbox{$-\overline{\psi}\gamma^\mu\gamma^5\psi\cdot A_{5\mu}$} in \mbox{Eq. (\ref{EMLagrange})} through the anomalous Ward identity.
}

\section{The case of constant magnetic field}\label{secConstB}
In this section we work out the electric current response in the presence of constant magnetic field and axial imbalance with arbitrary spacetime dependence. Physically, this approximation is meaningful if there is a separation between the scales of the perturbations in the axial imbalance field $\ChIm$ and the magnetic field: the latter has to vary on much larger space- and timescales. In RHIC, this is not the case for the whole lifetime of the QGP, but it can be a good approximation describing the initial state, when $\ve{B}$ is still large and relatively unchanged because of the conductivity of the medium. \REM{In Ref. \cite{JLmagnetic}, the interested reader can find a detailed analysis of the fluctuation pattern of the magnetic field in RHIC.} Due to the intense color fields, a region is formed where the axial imbalance is effectively nonzero. This region, however, is still affected by the fast gluonic dynamics, leaving the imbalance field to change fast as well.

\remC{As the time scale of the chiral charge creation/annihilation in RHIC may be comparable to or shorter than the thermal relaxation time, the hydrodynamic approximation for CME breaks down and one has to consider a space-time dependent chiral charge density. Without a first principle treatment of the nonperturbative dynamics of the chiral charge in QCD, a simple-minded assumption amounts to extend the constant chiral chemical potential to a space-time dependent one, which will be relied on in the rest of this paper. Under this assumption, the chiral chemical potential proxies the QCD dynamics of the axial imbalance and can be viewed as the temporal component of an axial-vector potential.} Keeping things simple we suppose $A=(0,\ve{A})$ and $A_5=(\actSign\ChIm,\ve{0})$, so there are no electric and axial magnetic fields present. It is straightforward to check that $\wt{\Gamma}^{0ij}_{AVV}(q_1=0,q_2)\equiv 0$ by using the Ward-identity to transform the integrand of the AVV vertex into a full derivative with respect to integration momentum. The finite mass and the PV-term then cancel each other out. The first contributing term in the small$-q_1$ expansion is $\frac{\partial \wt{\Gamma}^{0ij}_{AVV}}{\partial q_{1k}}$. Equivalently, one can plug the time-independent $\ve{A}(\ve{y})=\frac{1}{2}\ve{B}\times\ve{y}$ into \mbox{Eq. (\ref{responseCurrent})} to obtain the relation:
\begin{widetext}
\begin{align}
\avr{J^i}(x) =& \intlimB{^4q_2}{}{}\FTChIm(q_2)e^{iq_2\cdot x}\left[\actSign\frac{i}{2}\epsilon^{jlk}B^l\left.\frac{\partial \wt{\Gamma}^{0ij}_{AVV}(q_1,q_2)}{\partial q_{1k}}\right|_{q_1=0}\right] = 
\intlim{t'}{-\infty}{\infty}\intlimB{^3\ve{q}}{}{}\OLChIm(t',\ve{q})\overline{\sigma}^i_{A}(t'-t,\ve{q})e^{-i\ve{q}\cdot\ve{r}}.
\end{align}
\end{widetext}
where the kernel
\begin{align}
\overline{\sigma}^i_{A}(t,\ve{q})=\actSign\intlimB{q_0}{-\infty}{\infty}e^{iq_0t}\frac{i}{2}\epsilon^{jlk}B^l\left.\frac{\partial \wt{\Gamma}^{0ij}_{AVV}(q_1,q_2=q)}{\partial q_{1k}}\right|_{q_1=0}
\end{align}
is the CME conductivity in the mixed representation of spatial momentum and time, and an explicit formula of it will be derived below.
We shall omit the averaging sign $\avr{.}$ in what follows if we can without causing confusion. 
Before taking the derivative of \mbox{Eqs. (\ref{GAVVline1})-(\ref{GAVVline7})} with respect to $q_{1k}$, we note that the sum of \mbox{Eqs. (\ref{GAVVline1})}, (\ref{GAVVline2}), and (\ref{GAVVline3}) equals to the sum of \mbox{Eqs. (\ref{GAVVline4})}, (\ref{GAVVline5}), and (\ref{GAVVline6}), as can be demonstrated by transposing the matrices under the trace of the former employing the charge conjugation property $C(\gamma^\text{T})^\mu C^\dagger=-\gamma^\mu$ with $C=\gamma^2\gamma^0$ followed by transforming the integration momentum $p\to-p$. The transformation of the integration momentum is legitimate as long as the regulator terms kept in the scene. Consequently,
\begin{align}
\wt{\Gamma}^{\rho\mu\nu}_{AVV}(q_1,q_2)=\mathfrak{g}^{ij}_m(q_1,q_2)-\lim_{M\to\infty}\mathfrak{g}^{ij}_M(q_1,q_2),
\end{align}
where
\begin{widetext}
\rem{\begin{align}
\mathfrak{g}^{ij}_m(q_1=(0,\ve{q}_1),q_2=q) =
-e^2\int_p\text{tr}&\left\{ -\gamma^i G^A(p+q)\gamma^0\gamma^5 G^A(p-q_1) \gamma^jG^A(p)\cdot(1-2\wt{n}(p_0+q_{0})) +\right. \nn\\
& +\gamma^i G^R(p+q)\gamma^0\gamma^5 G^R(p) \gamma^jG^R(p-q_1)\cdot(1-2\wt{n}(p_0)) + \nn\\
& \left. +\gamma^i G^R(p+q)\gamma^0\gamma^5 G^A(p) \gamma^jG^A(p-q_1)\cdot(2\wt{n}(p_0)-2\wt{n}(p_0+q_{0})) \right.\}. \nn
\end{align}}
Here we have replaced $G^C(p)$ with ${\textstyle(1-2\wt{n}(p_0))(G^R(p)-G^A(p))}$. It follows that
\begin{align}
\frac{\partial}{\partial q_{1k}}\mathfrak{g}^{ij}_m(q_1\rightarrow 0,q_2=q) = e^2\int_p\text{tr}&\left\{ -\gamma^i G^A(p+q)\gamma^0\gamma^5 G^A(p) \gamma^j\frac{\partial}{\partial p_k}G^A(p)(1-2\wt{n}(p_0+q_{0})) + \right. \label{gAVVconstBline1} \\
& +\gamma^i G^R(p+q)\gamma^0\gamma^5 G^R(p) \gamma^j\frac{\partial}{\partial p_k}G^R(p)(1-2\wt{n}(p_0)) + \label{gAVVconstBline2}\\
& \left. +\gamma^i G^R(p+q)\gamma^0\gamma^5 G^A(p) \gamma^j\frac{\partial}{\partial p_k}G^A(p)(2\wt{n}(p_0)-2\wt{n}(p_0+q_{0})) \right\} \label{gAVVconstBline3}
\end{align}
\end{widetext}
and
\begin{align}
\overline{\sigma}^i_{A}(t,\ve{q})=\intlimB{q_0}{-\infty}{\infty}e^{iq_0t}\frac{i}{2}\epsilon^{jlk}B^l
&\left[\left.\frac{\partial \mathfrak{g}^{ij}_{m=0}(q_1,q_2)}{\partial q_{1k}}\right|_{q_1=0,q_2=q} \right.\nn \\
&\left.-\lim_{M\to\infty}\left. \frac{\partial \mathfrak{g}^{ij}_{m=M}(q_1,q_2)}{\partial q_{1k}}\right|_{q_1=0,q_2=q}\right]. \label{kernel}
\end{align}

The first step of evaluation is to determine the tensor structure of the expression. This in general would be tedious because there are three propagators left under the trace. However, we are interested in antisymmetric combinations in $jk$ only, because of the contraction with the magnetic field. The details of the trace calculation can be found in Appendix \ref{app:trace}. The resulting expression turns out to remain fairly compact:
\begin{widetext}
\begin{align}
\frac{\partial \mathfrak{g}^{ij}_m}{\partial q_{1k}}\epsilon^{ljk}B_l =& -16\pi e^2\int_p (1-2\wt{n}(p_0))\text{sgn}(p_0) \frac{1}{2|\ve{p}|}\times \nonumber \\
\times & \left[\left(B^i\left(m^2-(p_0+q_{0})p_0\right)-p^i\ve{B}\cdot(\ve{p}+\ve{q})\right) \frac{1}{1+\frac{\ve{p}\cdot\ve{q}}{\ve{p}^2}} \frac{\partial}{\partial |\ve{p}|} \left(\frac{1}{(p_0+q_{0}+i0^+)^2-(\ve{p}+\ve{q})^2-m^2}\right)\delta\left(p_0^2-\ve{p}^2-m^2\right) +\right. \nonumber \\
& \left. +\left(B^i\left(m^2-p_0(p_0+q_{0})\right)-(p^i+q^i)\ve{B}\cdot\ve{p}\right)\frac{1}{(p_0+q_{0}+i0^+)^2-(\ve{p}+\ve{q})^2-m^2}\frac{\partial}{\partial |\ve{p}|} \delta\left(p_0^2-\ve{p}^2-m^2\right) \right]. \label{gAVVconstBderivation1}
\end{align}
\end{widetext}
Here we regrouped the terms from the three propagator-products $AAA$, $RRR$ and $RAA$. This way it is possible to deal with the higher-order poles by recasting part of the expression as a derivative of either $\rho$ or $G^R$ --- more details of the calculation can be found in Appendix \ref{app:constB}. It proves to be useful to separate the components of $\ve{B}$ parallel to $\ve{q}$: $\ve{B}_\parallel=(\ve{B}\cdot\widehat{\ve{q}})\widehat{\ve{q}}$, and perpendicular to it: $\ve{B}_\perp=\ve{B}-\ve{B}_\parallel$. In this way part of the directional integration can be performed, leaving us with the following expression, the azimuthal integration still left to be done: 
\begin{widetext}
\begin{align}
 & \frac{\partial\mathfrak{g}^{ij}_m}{\partial q_{1k}}\epsilon^{ljk}B_l=\nonumber \\
=& -\frac{e^2}{\pi^2}\intlim{p_0}{-\infty}{\infty}\text{sgn}(p_0)(1-2\wt{n}(p_0))\intlim{p}{0}{\infty}\delta\left(p_0^2-p^2-m^2\right)\intlim{x}{-1}{1} \left\{ \frac{\partial}{\partial x}\frac{B_\parallel^i x(p_0q_{0}+qpx+(x^2+1)p^2) +B_\perp^i x\left(p_0q_{0}+\left(\frac{1-x^2}{2}+1\right)p^2\right)}{(p_0+q_{0}+i0^+)^2-p^2-q^2-2qpx-m^2} \right. + \nonumber \\
& \left. + B_\perp^i \frac{p^2}{(p_0+q_{0}+i0^+)^2-p^2-q^2-2qpx-m^2} \right\}. \label{constBrespMomspaveLine2}
\end{align}
\end{widetext}
Note that the first term in the above expression is a total derivative with respect to $x$. 

The contributions of the two terms in the integrand of \mbox{Eq. (\ref{kernel})} are calculated separately with the detailed steps laid out in Appendix \ref{app:constB}. For the massless term, we take the Fourier transform of \mbox{Eq. (\ref{constBrespMomspaveLine2})} with respect to $q_0$ first and calculate the rest of the integrals afterwards. All integrations can be carried out analytically for $m=0$ and we obtain that
\begin{widetext}
\begin{align}
\intlimB{q_0}{-\infty}{\infty} e^{iq_0t}\frac{\partial\mathfrak{g}^{ij}_{m=0}}{\partial q_{1k}}\epsilon^{ljk}B_l=\theta(-t)\frac{e^2}{\pi^2} (B^i-\widehat{q}^i(\ve{B}\cdot\widehat{\ve{q}})) t\frac{\partial}{\partial t}\left(\frac{\sin(qt)}{qt}\right)T\intlim{y}{0}{\infty}(1-2n_{FD}(y))\sin(2yTt).\label{chiral_term}
\end{align}
For the PV term, we scale the loop momentum $p$ by the regulator mass $m=M$ as $p=My$ and take the limit $M\to \infty$. The rest of the integrals can be calculated analytically with the 
result
\begin{align}
\intlimB{q_0}{-\infty}{\infty}e^{iq_0t}\frac{\partial\mathfrak{g}^{ij}_{M\rightarrow\infty}}{\partial q_{1k}}\epsilon^{ljk}B_l = & \frac{e^2}{\pi^2}\left\{-B^i\delta(t) +\theta(-t)\left[\left(B_\parallel^i+\frac{B_\perp^i}{2}\right)\frac{\partial^2}{\partial t^2}\left(\frac{\sin(qt)}{q}\right) +\frac{B_\perp^i}{2}\frac{\partial}{\partial t}\left(\frac{\sin(qt)}{qt}\right)\right] \right\}.
\label{PV_term}
\end{align}

By combining the two, now we are equipped with the mixed representation of the CME conductivity in case of constant, homogeneous magnetic field. 

\begin{align}
\overline{\sigma}_{A}^i(t,\ve{q}) =& \actSign\frac{1}{2} \intlimB{q_0}{-\infty}{\infty}e^{iq_0t}\left(-\frac{\partial\mathfrak{g}^{ij}_{M\rightarrow\infty}}{\partial q_{1k}}+\frac{\partial\mathfrak{g}^{ij}_{m=0}}{\partial q_{1k}}\right)\epsilon^{ljk}B_l = \nn\\
=& \actSign\frac{e^2}{2\pi^2}\left\{B^i\delta(t)+\frac{\theta(-t)}{2}\left[q\sin(qt)\left(B^i+\widehat{q}^i(\ve{B}\cdot\widehat{\ve{q}})\right)-\frac{\partial}{\partial t}\left(\frac{\sin (qt)}{qt}\right)f(tT)\left(B^i-\widehat{q}^i(\ve{B}\cdot\widehat{\ve{q}})\right) \right]\right\}, \label{constBrespFunc}
\end{align}
where $f(x)$ comes from the Fermi-Dirac distribution:

\begin{align}
f(x) =& 4x\intlim{y}{0}{\infty}n_{FD}(y)\sin(2yx) = 1-\frac{2\pi x}{\sinh(2\pi x)} \longrightarrow \left\{\begin{array}{lll} 0, & & x\rightarrow 0 \\ 1, & & x\rightarrow\infty \end{array}\right. \label{funcSmallF}
\end{align}
\rem{
Performing a Fourier transformation on \mbox{Eq. (\ref{constBrespFunc})} with respect to time, one obtains the frequency-momentum representation of the response function
\begin{align}
\wt{\sigma}_{A}^i(\omega,\ve{q}) =&\intlim{t}{-\infty}{\infty}e^{-i\omega t}\overline{\sigma}_{A}^i(t,\ve{q})
= \actSign\frac{e^2}{4\pi^2}\left\{2B^i+\frac{q^2}{\omega^2-q^2}[B^i+\widehat{q}^i(\ve{B}\cdot\widehat{\ve{q}})] +\right.\nn\\
&\left. +\intlim{p}{0}{\infty}\,n_{FD}(p/T)\left[\frac{1}{q}\ln\frac{\omega^2-(2p-q)^2}{\omega^2-(2p+q)^2} 	-2\frac{2p-q}{\omega^2-(2p-q)^2}-2\frac{2p+q}{\omega^2-(2p+q)^2}\right][B^i-\widehat{q}^i(\ve{B}\cdot\widehat{\ve{q}})]\right\}, \label{constBrespFuncF}
\end{align}
}
\end{widetext}
\rem{
where the frequency $\omega$ carries an infinitesimal positive imaginary part and the integral representation of $f(tT)$, \mbox{Eq. (\ref{funcSmallF})} is employed. The frequency-momentum representation of the electric current reads
\begin{equation}
\wt{J}^i(\omega,\ve{q})=\wt{\sigma}_{A}^i(\omega,\ve{q})\FTChIm(\omega,\ve{q}).
\end{equation}
It follows from the continuity equation $q\cdot\wt{J}=0$ that
\begin{equation}
\wt{n}(\omega,\ve{q})=\frac{\ve{q}\cdot\wt{\ve{J}}(\omega,\ve{q})}{\omega}=\actSign\frac{e^2}{2\pi^2}\FTChIm\frac{\omega}{\omega^2-q^2}\ve{q}\cdot\ve{B},
\end{equation}
where $\wt{n}$ is the Fourier transform of the charge density $n$. Interestingly, the resulting expression is temperature independent.

The above expressions in \mbox{Eqs. (\ref{constBrespFunc})} and (\ref{constBrespFuncF}) are the main original contribution of this paper. Although fairly complicated when being convoluted with a profile of $\mu_5$, these expressions are still suitable to investigate the charge transport in special situations as we will see in the next section. About to be proven practical as it is, one might find the coordinate representation of the response function more useful in other cases. For more details on that, see Appendix \ref{app:coordResp}.
}
\subsection{Limiting cases}\label{secConstBLimits}
In order to gain some insights into the expression in \mbox{Eq. (\ref{constBrespFunc})}, let us first analyze its behavior in two limiting cases.

A time-independent $\ChIm$ will render the conductivity in the zero frequency limit, which is equivalent to the integral of $\overline{\sigma}_A^i(t,\ve{q})$ with respect to its time-argument: 
\begin{widetext}
\begin{align}
\intlim{t}{-\infty}{\infty}\overline{\sigma}^i_{A} =& \actSign\frac{e^2}{2\pi^2}\left[ B^i +\left(B^i +\widehat{q}^i(\ve{B}\cdot\widehat{\ve{q}})\right)\frac{1}{2}\intlim{t}{-\infty}{0}q\sin qt -\left(B^i -\widehat{q}^i(\ve{B}\cdot\widehat{\ve{q}})\right)\frac{1}{2}\intlim{t}{-\infty}{0}\frac{\partial}{\partial t}\left(\frac{\sin qt}{qt}\right)f(tT)\right] = \nn\\
=& \actSign\frac{e^2}{4\pi^2}\left(B^i -\widehat{q}^i(\ve{B}\cdot\widehat{\ve{q}})\right)\left[1-\intlim{\tau}{-\infty}{0}\frac{\partial}{\partial\tau}\left(\frac{\sin\tau}{\tau}\right)f\left(\frac{\tau T}{q}\right)\right] \overset{q\rightarrow 0}{\longrightarrow} 0^i. \label{time_integrated}  
\end{align}
\end{widetext}
In the static but inhomogeneous limit ($\ve{q}\neq \ve{0}$) we see that the conductivity is perpendicular to $\ve{q}$. On one hand this means local charge conservation is fulfilled. On the other hand it shows that the current has a dipolelike structure, which has consequences regarding the long-time behavior of the charge transport, as we shall see soon.

To approach the limit of constant $\ChIm$ by sending $q\rightarrow 0$ one observes that the current vanishes, since $f(x\rightarrow\infty)=1$. We note here that this limiting behavior was already reported in Ref. \cite{WHR2017} and some aspects were also discussed in Ref. \cite{HLR2011}. It is not that surprising that the electric current vanishes for constant axial imbalance. The nonexistence of the CME at equilibrium was reported by several authors, see for example Refs. \citeNoEqCME. In the context of Weyl semimetals, where it is possible to prepare the system in such a way that the introduction of a chiral chemical potential makes sense, there is a consensus that in equilibrium the CME current vanishes --- even for nonzero $\ChIm$, see Refs. \cite{Shovkovy1,YamamotoBlochT}.

When the static limit is taken first, the UV-originated anomaly contribution is canceled by the following term: $\textstyle ie^2\int_p(-2\wt{n}'(p_0))\text{tr}\left\{\gamma^5\gamma^iG^A(p)\gamma^j\frac{\partial G^A(p)}{\partial p_k}\right\}$. Although we do not go into the details here, one can show that interactions will not change this expression, see for example Refs. \citeNoCorr. So the vanishing of the conductivity in the mentioned limit is a general result.

For homogeneous $\ChIm$ configurations, i.e., $\ve{q}\rightarrow 0$ only the $\frac{e^2}{2\pi^2}\ve{B}\delta(t)$ term in \mbox{Eq. (\ref{constBrespFunc})} survives. This term in the end provides the usual homogeneous current parallel to $\ve{B}$, tied to the anomaly. 
Let us expand \mbox{Eq. (\ref{constBrespFunc})} around to the first nontrivial order in $q$ to learn what happens if the system is pushed away from homogeneity:

\begin{align}
\delta\ve{J}(t,\ve{q})\equiv&\ve{J}(t,\ve{q}) -\frac{e^2}{2\pi^2}\overline{A}_{5,0}(t,\ve{q})\ve{B} \approx \nn \\
\approx \actSign\frac{e^2}{4\pi^2}\intlim{\tau}{-\infty}{0}& \OLChIm(t+\tau,\ve{q})q^2\tau\left[\left(1+{\textstyle\frac{1}{3}}f(\tau T)\right)\ve{B} +\right. \\
&\left. \left(1-{\textstyle\frac{1}{3}}f(\tau T)\right)(\ve{B}\cdot\widehat{\ve{q}})\widehat{\ve{q}}\right] +\mathcal{O}(q^4).
\end{align}
The above expression is still too complicated to arrive at a compact analytic expression. We now will assume that there is a clear separation between the internal timescale and temperature, namely, send $T$ either to 0 or $\infty$. In both cases we end up with the following expression:

\begin{align}
\delta\ve{J}(t,\ve{q}) \approx & \actSign\frac{e^2}{4\pi^2}\left(C_1q^2\ve{B} +C_2(\ve{B}\cdot\ve{q})\ve{q}\right)
\intlim{\tau}{-\infty}{0}\OLChIm(t+\tau,\ve{q})\tau
\end{align}
with the constants $\textstyle\left\{C_1,C_2\right\}$ being either  $\textstyle\left\{1,1\right\}$ for $T=0$ or $\textstyle\left\{\frac{4}{3},\frac{2}{3}\right\}$ for $T\rightarrow\infty$. After Fourier transform one arrives at the differential equation below:

\begin{align}
\partial_t^2\delta\ve{J}(t,\ve{r}) =& \actSign\frac{e^2}{4\pi^2}\left(C_1\ve{B}\nabla^2_\ve{r} +C_2(\ve{B}\cdot\nabla_\ve{r})\nabla_\ve{r} \right)\ChIm(t,\ve{r}). \label{hydroGradCorr} 
\end{align}
For homogeneous $\ChIm$ the right-hand side (RHS) of \mbox{Eq. (\ref{hydroGradCorr})} is zero, leading no deviation from the CME current in \mbox{Eq. (\ref{CME1})}. On the other hand, inhomogeneity can add to the dynamics of the electric current. One should, however, keep in mind that the short-time behavior of the corrections in small $\ve{q}$ provided by our weak-coupling calculation might be significantly modified at strong coupling \cite{LublinskyHolo1}.

Straightforward analysis shows that $\partial_t^2\nabla\times\delta\ve{J}$ is not zero: \remC{Taking the curl of \mbox{Eq. (\ref{constBrespFunc})} it has contributions from the temperature dependent part of the expression as well -- in contrast with the behavior of $\nabla\cdot\ve{J}$. In terms of the long-wavelength approximation of \mbox{Eq. (\ref{hydroGradCorr})} it is the $C_1$--term with nonzero curl.} This means that inhomogeneous $\ChIm$ can, even without sizable electric field or spatial dependence of $\ve{B}$, alter the vorticity of the current field. Depending on the dynamics of $\ChIm$ this might cause instabilities in a laminar charge flow, leading to the formation of vortices. Other studies indicate various instabilities in case of the chiral plasma is affected by dynamical EM fields, see for example Refs. \cite{CSeldinExample,TuchinHICdynamicEM,chiralHydroInst} or the recent first principle study of Ref. \cite{chiralPlasmaInst} simulating QED plasma. Further theoretical investigation is needed by taking the feedback of axial charge and EM fields into account. We will address this question in a future publication. 

\section{CME contribution to the charge asymmetry}\label{secChargeAsymm}
\remC{
The main goal of this section is to gain some insights on the implications of our results above for possible observables of CME in heavy ion collisions. In these collisions, the magnetic field points along the out-of-plane direction, i.e. perpendicular to the reaction plane. The CME current would then transport positive/negative charges in opposite direction across this reaction plane, and eventually leads to a charge dipole distribution. This can be measured through charge asymmetry in hadrons' azimuthal correlations. The key to this observable is the amount of electric charges being transported across the reaction plane. While we are not simulating a heavy ion collision here, we try to obtain some insights into this problem by analyzing the amount of transported charge by the CME current through a transverse area on the plane perpendicular to the magnetic field for certain chiral charge distribution patterns motivated by heavy ion collisions. In particular we focus on the long time behavior of the net charge asymmetry across such transverse area and discuss the implications.
}

\subsection{Long-time behavior after quench}
\begin{figure}[!h]
		\includegraphics[width=0.85\linewidth]{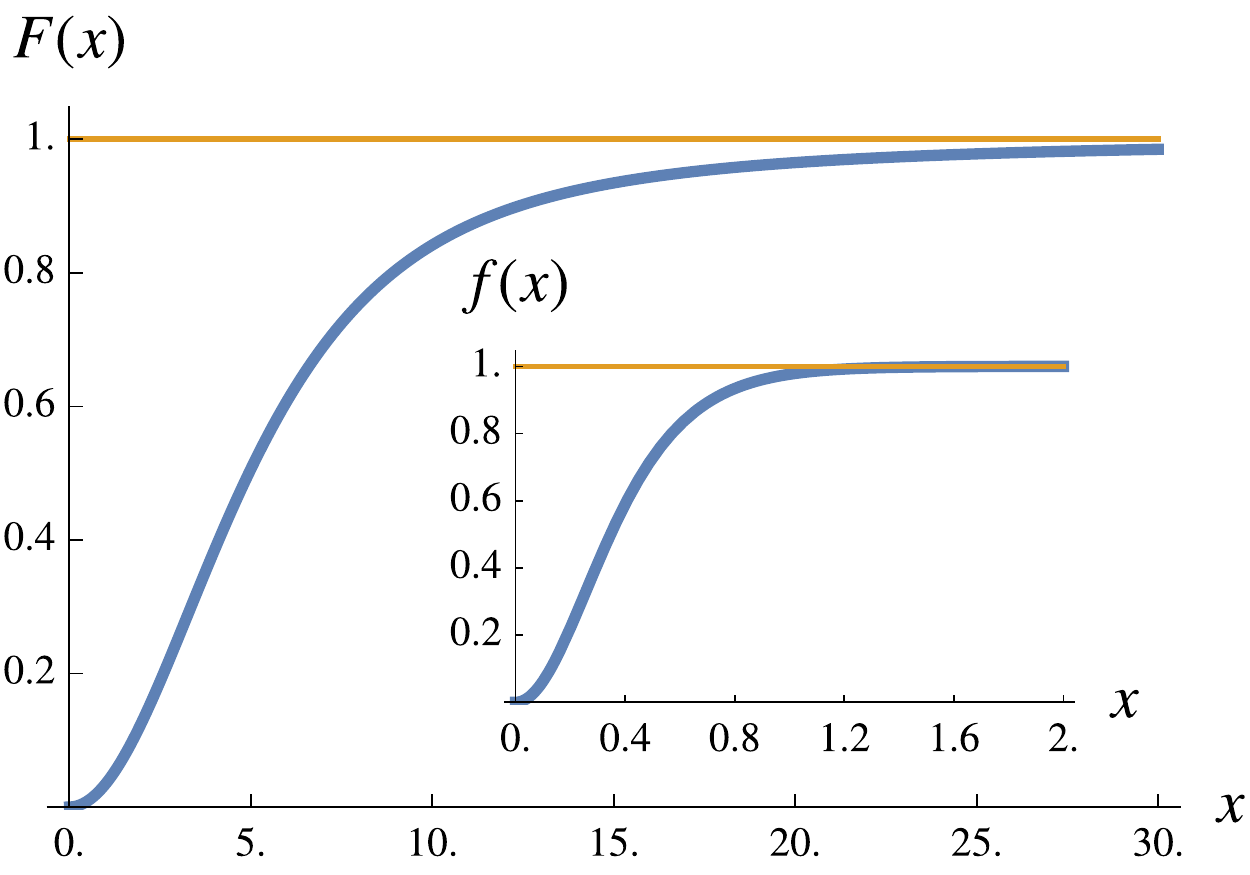} 
	\caption{Functions characterizing the response function. $f$ is defined in \mbox{Eq. (\ref{funcSmallF})}, whereas $F$ is derived from $f$ in order to give the long-time behavior of the response, see \mbox{Eq. (\ref{funcCapitalF})}.}\label{fig:functions}
\end{figure}

First we consider the scenario when there is a sudden onset of the axial imbalance, corresponding to $\OLChIm(t,\ve{q})=\theta(t)\FTChIm(\ve{q})$. \remB{Although at first such a perturbation might seem to be out of reach for the WFA, we note that for weak enough external fields the expressions in \mbox{Eqs. (\ref{Jlinres}-\ref{J5linres})} are justified --, \textit{any} space-time dependence of $A_{5,0}$ is fine, since the response function takes all orders of gradients into account.} The electric current is given by this expression:
\begin{align}
\overline{J}^i(t,\ve{q}) =& \actSign\frac{e^2}{2\pi^2}\FTChIm(\ve{q}) \left[B^i -\frac{1}{2}(B^i+\widehat{q}^i(\ve{B}\cdot\widehat{\ve{q}}))\intlim{\tau}{0}{t}q\sin q\tau +\right. \nn\\
&\left. +\frac{1}{2}(B^i-\widehat{q}^i(\ve{B}\cdot\widehat{\ve{q}}))\intlim{\tau}{0}{t}\frac{\partial}{\partial\tau}\left(\frac{\sin q\tau}{q\tau}\right)f(\tau T)\right],
\end{align}
which for long times simplifies further to

\begin{align}
\overline{J}^i(t,\ve{q}) \overset{t\rightarrow\infty}{\longrightarrow} & \actSign\frac{e^2}{2\pi^2}\frac{\FTChIm(\ve{q})}{2}\left(B^i-\frac{q^i(\ve{B}\cdot\ve{q})}{q^2}\right)F(q/T), \label{CMEcurrLongTime}
\end{align} 
where

\begin{align}
F(x) =& 1+\intlim{y}{0}{\infty}\frac{\partial}{\partial y}\left(\frac{\sin xy}{xy}\right)f(y) =1-\intlim{y}{0}{\infty}\frac{\sin xy}{xy}f'(y) \nn \\
& \longrightarrow\left\{\begin{array}{ccc} 0, & & x\rightarrow 0 \\ 1, & & x\rightarrow\infty \end{array}\right. \label{funcCapitalF}
\end{align}
The function $F(x)$ as plotted in Fig. \ref{fig:functions} vanishes for small $x$ and approaches 1 monotonically for large $x$. This results in $F$ acting as an infrared cutoff when the Fourier transform is performed to get the coordinate-space expression for \mbox{$\ve{J}(t\rightarrow\infty,\ve{r})$}. \rem{Note that the current is divergence free, \mbox{$\nabla_\ve{r}\cdot\ve{J}(t\rightarrow\infty,\ve{r})=0$}, as can be seen form the $\ve{q}$-dependence in \mbox{Eq. (\ref{CMEcurrLongTime})}, so it can be expressed as the curl of another vector field. Such a current is usually referred to as a magnetization current.} The suppression of the small $\ve{q}-$domain makes $\ve{J}$ to be localized in a region with size controlled by $1/T$. In the limiting case $\ve{q}\rightarrow 0$ the current is zero --- as expected for $\ChIm$ when the homogeneity limit is taken after the static limit. The result is the same if for some reason $T$ supersedes any (inverse) spatial scales, since $T\rightarrow\infty$ renders $\overline{J}^i$ to be zero through $F(0)=0$ again. For the other extreme, $T\rightarrow 0$, $J^i$ reveals a dipole pattern:

\begin{align}
\left.\ve{J}(t\rightarrow\infty,\ve{r})\right|_{T=0} =& \actSign\frac{e^2}{16\pi^3}\intlim{^3\ve{r}'}{}{}\frac{-1}{|\ve{r}-\ve{r}'|}\nabla_{\ve{r}'}\times (\ve{B}\times \nabla_{\ve{r}'}) \ChIm(\ve{r}')= \nn\\
=& \actSign\frac{e^2}{16\pi^3}\intlim{^3\ve{r}'}{}{}\frac{\ve{B}-\frac{3(\ve{B}\cdot\ve{r}')\ve{r}'}{(r')^2}}{(r')^3}\ChIm(\ve{r}-\ve{r}') =\nn\\
=& \actSign \frac{e^2}{16\pi^3}\nabla_{\ve{r}}\times(\ve{B}\times\nabla_{\ve{r}})\intlim{^3\ve{r}'}{}{}\frac{-\ChIm(\ve{r}-\ve{r}')}{r'}.
\end{align}
One can show that this current dipole transports zero charge in total through a large enough surface perpendicular to the direction of the magnetic field. In general, this is the consequence of the structure $\ve{B}-\widehat{\ve{q}}(\ve{B}\cdot\widehat{\ve{q}})$. However, a well-localized source is enough to explain what happens: the ``current field lines'' are closed because of $\nabla_{\ve{r}}\cdot\ve{J}(t\rightarrow\infty,\ve{r})=0$ so any of them travels through twice on a large enough surface. We can easily show this with a pointlike source using the previous dipole formula and integrating over the surface $S\perp\ve{B}$:

\begin{align}
&\intlim{^2\ve{r}}{S}{} \left.\widehat{\ve{B}}\cdot \ve{J}(t\rightarrow\infty,\ve{r})\right|_{T=0} \propto \intlim{^2\ve{r}}{S}{}\widehat{\ve{B}}\cdot\nabla\times\frac{\ve{B}\times\ve{r}}{r^3} = \nn\\
&= \oint\limits_{\partial S}\mathrm{d}\ve{l}\cdot \frac{\ve{B}\times\ve{r}}{r^3} \propto \frac{1}{r} \overset{r\rightarrow \infty}{\longrightarrow} 0,
\end{align}
where we used Stokes's theorem and in the last step we recognized that $\mathrm{d}\ve{l}$ and $\ve{B}\times\ve{r}$ are parallel and both are proportional to $r$. Since this observation is based on the long-time behavior of the current, {only after sufficiently long time it is true that the net transported charge does not change.}

An interesting side-note can be made at this point. Let us further analyze the long-time behavior of the current by integrating it over a spherical region $S_R$ ($R$ being the radius):

\begin{align}
&V\cdot \overline{\ve{J}} := \intlim{^3\ve{r}}{S_R}{}\ve{J}(t\rightarrow\infty,\ve{r}) =\nn\\
&= \actSign\frac{e^2}{2\pi^2}\frac{1}{2}\intlimB{^3\ve{q}}{}{}\frac{\sin qR - qR\cos qR}{q^3}F(q/T)\FTChIm(\ve{q})\left(\ve{B}-\frac{\ve{q}(\ve{B}\cdot\ve{q})}{q^2}\right).
\end{align}
with the volume $V$ of $S_R$ and $\ve{J}(t\rightarrow\infty,\ve{r})$ as the coordinate representation of \mbox{Eq. (\ref{CMEcurrLongTime})}. Now we consider a source of $\ChIm$ centered in space around the origin, whose Fourier transform is $\FTChIm(\ve{q})=V\cdot\ChIm$. After some calculation whose details can be found in Appendix \ref{app:avrJquenched} we arrive at:

\begin{align}
\overline{\ve{J}} =&\actSign\frac{e^2}{2\pi^2}\ChIm\frac{1}{\pi}\frac{2}{3}\ve{B}\intlim{Q}{0}{\infty}F(q/(RT))\frac{\sin Q -Q\cos Q}{Q} = \nn\\
=& \actSign\frac{e^2}{2\pi^2}\ChIm\ve{B}\frac{1-f(RT)}{3}. \label{constBrespQuenchedLTAvr}
\end{align}
This expression does not carry the dipole structure anymore. Instead, there is a suppression factor of $(1-f(RT))/3$: for zero temperature or when the spatial averaging is done within an asymptotically small sphere the result is the $1/3$ of the anomaly ruled current. Sending either $R$ or $T$ to large values, however, renders $\overline{\ve{J}}$ to zero. Generally, the expression is monotonically interpolates between this two limiting cases depending on the relative value of $R$ and $T$.

\subsection{Long-time charge transport parallel to $\ve{B}$}
Returning to the question of transported electric charge, one can argue that its vanishing behavior for long times is a generic feature. This is closely related to the fact that the local charge conservation $\partial\cdot J=0$ is an essential property of the system. Due to the CME there is electric current in the direction of the magnetic field. The system does not have any boundary, so it is quite natural that the charge flows back somewhere: since the current tends to be parallel to $\ve{B}$ in the presence of chiral imbalance, the back-flow happens away from these regions, where $\ChIm$ vanishes. We have already seen this dipole structure at work in the previous subsection. Therefore, taking a large enough surface perpendicular to $\ve{B}$, we expect that the net charge through this surface tends to zero as time passes.

\remB{Throughout this subsection we show the vanishing of the long-time transported charge parallel to the magnetic field. Firstly, by using the weak coupling result in \mbox{Eq. (\ref{CMEcurrLongTime})}, then to generalize our statement beyond perturbation theory, the local charge conservation is invoked -- in terms of the Ward-identity for the AVV vertex function, \mbox{Eq. (\ref{gradJ})}. }In order to put the argument onto more general grounds we analyze the following quantity:

\begin{align}
\Delta Q_S :=& \intlim{t}{-\infty}{\infty}\intlim{^2\ve{r}}{S}{}\widehat{\ve{B}}\cdot\ve{J}(t,\ve{r}), \label{DeltaQInf}
\end{align}
where the surface $S$ is the plane with the normal vector $\widehat{\ve{B}}$. Utilizing the conductivity relation in \mbox{Eq. (\ref{constBrespFunc})} we can write
\begin{align}
\Delta Q_S =& \intlimB{^3\ve{q}}{}{}\intlim{^2\ve{r}}{S}{}\widehat{B}_i \wt{\sigma}_A^i(q_0=0,\ve{q})\FTChIm(q_0=0,\ve{q})e^{-i\ve{q}\cdot\ve{r}} =\nn\\
=& \actSign\frac{e^2B}{4\pi^2}\intlimB{^3\ve{q}}{}{}\intlim{^2\ve{r}}{S}{} \left(1 -(\widehat{\ve{B}}\cdot\widehat{\ve{q}})^2\right)\times\nn\\
&\underbrace{\left[1-\intlim{\tau}{-\infty}{0}\frac{\partial}{\partial\tau}\left(\frac{\sin\tau}{\tau}\right)f\left(\frac{\tau T}{q}\right)\right]}_{=F(q/T)}\FTChIm(0,\ve{q})e^{-i\ve{q}\cdot\ve{r}}.
\end{align}
Integrating over the surface $S$ when its size is large enough, the components of $\ve{q}$ parallel to $S$ are forced to vanish and only the component parallel to $\ve{B}$ survives:

\remC{\begin{align}
\Delta Q =& \lim\limits_{\text{area of }S\rightarrow\infty}\Delta Q_S =\actSign\frac{e^2B}{4\pi^2} \intlimB{q_\parallel}{-\infty}{\infty}F\left(\sqrt{\smash[b]{q_\parallel^2+q_\perp^2}}/T\right)\times \nn \\
& \left.\left(1-\frac{q_\parallel^2}{q_\parallel^2+q_\perp^2}\right)\FTChIm(0,q_\parallel\widehat{\ve{B}}+\ve{q}_\perp)\right|_{\ve{q}_\perp=0} = 0. \label{DeltaQInfZero}
\end{align}}
In conclusion, the net charge transported by the CME current for any chiral imbalance in constant magnetic field \textit{vanishes for long enough time when it is measured through a large enough plane perpendicular to $\ve{B}$}. 

\REM{
It is worth pointing out that the above statement is actually the consequence of a more general feature of the vertex function.
\rem{We can even proceed beyond the weak coupling approximation: summing up the transported charge for a large enough plane for long enough time one has to get a vanishing net result because of local charge conservation!} Taking an infinitely large plane with the normal vector $\widehat{\ve{n}}$ the transported charge $\Delta Q$ can be written most generally as follows:
\begin{align}
\Delta Q=& \intlimB{^4q}{}{}\intlimB{q_\parallel}{-\infty}{\infty}\widehat{n}_i\wt{\Gamma}^{0ij}_{AVV}(q,q')\wt{A}_j(q)\wt{A}_{5,0}(q'),
\end{align}
with $q=(q_0,\,\ve{q})$ and $q'=(-q_0,\,-\ve{q}+q_\parallel\widehat{\ve{n}})$. Now we recognize that because of $q+q'=(0,\,q_\parallel\widehat{\ve{n}})$, the combination in the integrand of the above equation can be recasted as the first identity in \mbox{Eq. (\ref{gradJ})}: $q_\parallel\widehat{n}_i\wt{\Gamma}^{0ij}_{AVV}(q,q')=(q+q')_\mu\wt{\Gamma}^{0\mu j}_{AVV}(q,q')\equiv 0$, rendering indeed $\Delta Q$ to vanish. Therefore the previous statement on the vanishing of the long-time transported charge $\Delta Q$ is generalized beyond the weak-coupling approximation\rem{: it is a consequence of the Ward-identity, i.e. local charge conservation, which is valid in \textit{all} orders of the perturbation theory.}}

\subsection{Interplay of many scales in charge asymmetry}
The transported charge shows ambiguity when a homogeneous time-dependent source is considered: performing the $\ve{q}-$integration puts the conductivity in the homogeneity limit --- ruled by the anomaly --- so one gets the standard CME current. On the other hand, performing the $t-$integration first is equivalent by taking the static limit first. Regardless of the $\ve{q}-$dependence, the transported charge is zero in this case because of the dipolar structure of the integrated conductivity, as we mentioned above. The ambiguity boils down to the fact that we have to deal with the different orders of limits --- mentioned in Sec. \ref{secConstBLimits} --- of $\ve{q}\rightarrow 0$ and $q_0\rightarrow 0$ when computing the transported charge of a homogeneous source. So it seems, $\Delta Q$ for constant $\ChIm$ is ill defined.

The apparent contradiction can be resolved by modifying the definition of our observable by taking into account the timescale of observation, $t_\text{obs.}$:
\begin{align}
\Delta Q(t_\text{obs.}) =& \intlim{t}{-t_\text{obs.}/2}{t_\text{obs.}/2}\intlim{^2\ve{r}}{S\perp\ve{B}}{}\widehat{\ve{B}}\cdot\ve{J}(t,\ve{r}). \label{DeltaQfinite}
\end{align}

In order to give meaningful a physical interpretation, let us consider the following different situations. In the case when the observation time is way much longer than the lifetime of the source --- regardless of its spatial structure --- taking $q_0\rightarrow 0$ first is justified, so $\Delta Q$ vanishes. 

The opposite order of limits is a good approximation only if the source is homogeneous throughout the whole time-evolution --- in that case taking $\ve{q}\rightarrow 0$ first is justified and the charge transport is given by the CME expression in \mbox{Eq. (\ref{CME1})}. But this scenario is rather unphysical when the observation time is very long: eventually the source $\ChIm$ has to have boundaries in space and/or time. So to relax the ambiguity, one should abandon the infinite-time integration in $\Delta Q$ and integrate only over a finite period while $\ve{q}\rightarrow 0$ is a good approximation.

\begin{figure}[!h]
		\includegraphics[width=0.85\linewidth]{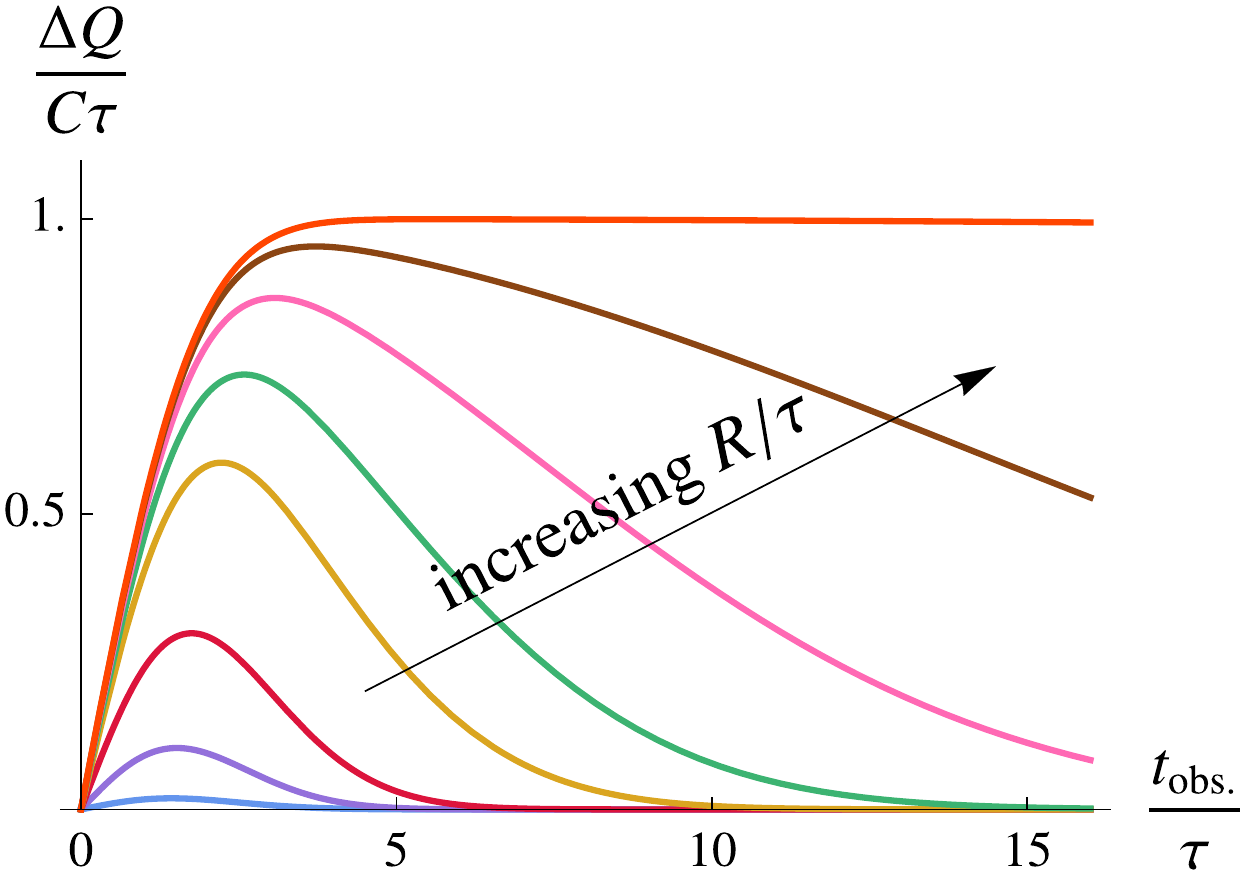} 
	\caption{Behavior of the transported charge $\Delta Q$ over the source lifetime $\tau$. $t_\text{obs.}-$dependence of $\frac{\Delta Q}{C\tau}$ for fixed $\tau$ and varying $R/\tau=\{$(0.2\textit{ (blue)}, 0.5\textit{ (purple)}, 1.0\textit{ (red)}, 2.0\textit{ (golden)}, 3.0\textit{ (green)}, 5.0\textit{ (pink)}, 10.0\textit{ (brown)}, 100.0\textit{ (orange)}$\}$: lower value leads the response to vanish faster and in general reach smaller maximum value. For large enough $R$ we see the homogeneous limit behavior. $\textstyle C=-\mu_0\frac{e^2}{2\pi^2}BA$}\label{fig:deltaQvarR}
\end{figure}

Let us demonstrate this phenomenon by using a simple toy-model within which the current is induced by the axial imbalance profile:
\remC{\begin{align}
\ChIm(t,r) =& \frac{\mu_0}{\sqrt{\pi}}e^{-\frac{t^2}{\tau^2}-\frac{r^2}{R^2}}, \label{Mu5Impulse}
\end{align}}
characterized by its spatial size $R$ and its lifetime $\tau$. 
The constant $\mu_0$ is to set the total axial charge generated by the source throughout its time-evolution. \REM{\rem{One can think of this profile as a very crude model of a ``fireball'' with Gaussian axial charge density. However, we stress here that} the above ansatz is meant only to show how the different scales interplay. One needs to investigate further the actual underlying cause of the axial imbalance in order to give a realistic description of the system at hand.} 
The infinite size of the plane $S$ again simplifies the result, making the finite temperature contribution vanish. The integration can be carried out analytically with the technical details and the lengthy formula of $\Delta Q$ presented in Appendix \ref{app:ChargeTranspMu5impulse}. As shown there, the scaled charge transport $\textstyle \frac{\Delta Q}{C \tau}$ with $\textstyle C=-\mu_0\frac{e^2}{2\pi^2}BA$ and {$A=\pi R^2$ --- being the effective size of the source ---} is a function of the dimensionless observation time $\textstyle \frac{t_\text{obs.}}{\tau}$ and the dimensionless extension of the axial imbalance $\textstyle \rho=\frac{R}{\tau}$. In Fig. \ref{fig:deltaQvarR}, we plotted $\textstyle \frac{\Delta Q}{C \tau}$ versus $\textstyle \frac{t_\text{obs.}}{\tau}$ for different values of $\rho$. By increasing $R$ with fixed $\tau$ and the observation time $t_\text{obs.}\gg\tau$, we can see the transition from the case of a well-localized source, $\textstyle\rho\ll 1$ which leads to vanishing $\Delta Q$, to the homogeneous source limit where $\textstyle \rho\gg 1$ and $\textstyle\Delta Q =-\mu_0\frac{e^2}{2\pi^2}BA\tau$.  So in case of an axial imbalance source with a very large but finite spatial size, one has to wait long enough for the transported charge to disappear. If the corresponding timescale, characteristic to the source, is much larger compared to the observation time, the charge transport is effectively described by the expression of \mbox{Eq. (\ref{CME})}. As the difference between different orders of limit is robust against higher order corrections, the transition described here from a localized axial imbalance to an extensive axial imbalance remains qualitatively valid for all orders of perturbation theory.

We conclude this section by emphasizing again that in order to say anything conclusive about a physically realistic situation, one needs to know the actual evolution of the axial imbalance. As we have seen above, depending on the various scales of the nonequilibrium system, the outcome can be vastly different, interpolating between the two limiting behaviors of the CME conductivity discussed in Sec. \ref{secConstBLimits}. \remC{This eventually leads to the behavior of the transported charge interpolating between a fast-disappearing transient and a long-lasting charge asymmetry.} All of these concerns point us to the need of dynamically more detailed and realistic simulations of the anomalous transport in QGP, in order to capture a real CME signal. \remC{Our findings -- although not directly applicable in RHIC phenomenology, -- highlight the sensitivity of the CME signal to the details of the axial imbalance. }

\section{Discussion and outlook}\label{secConclusions}

In this paper we analyzed the chiral magnetic current in constant magnetic field but with an arbitrary axial charge imbalance. In weakly coupled QED we derived the explicit form of the real-time response function in \mbox{Eq. (\ref{constBrespFunc})}, which interpolates between the anomaly ruled CME current like in \mbox{Eq. (\ref{CME})} --- with opposite sign compared to that --- and zero current at equilibrium, depending on the spatial pattern and time-dependence of the axial imbalance field $A_{5,0}=-\mu_5$. Then we explored the consequences of \mbox{Eq. (\ref{constBrespFunc})} for different spatio-temporal patterns of $\ChIm$. The observation that the static chiral imbalance results zero response current in the homogeneous limit shows the inherent nonequilibrium nature of the CME, as was already pointed out by others \cite{Rubakov,WHR2017,FukushimaNonstatCME,Shovkovy1,YamamotoBlochT} and this property is robust against higher order corrections \citeNoCorr. \remB{The reason behind the opposite sign of the anomaly ruled limit of \mbox{Eq. (\ref{CME1})} is the UV regularization needed to ensure local electric charge conservation even for an arbitrary $\mu_5$. As a consequence, the result of the triangle diagram is complemented by an additional contribution deriving from the UV sector. A Chern-Simons term in the effective action has the equivalent effect.} 

Coming to the phenomenological implications, we computed the electric current through the plane perpendicular to the constant magnetic field. For a localized axial imbalance, we found that the total electric charge transported through the plane over a long time vanishes because of the dipolar spatial structure of the time-integrated CME conductivity $\sigma_A^i(t,\ve{r})$, \mbox{Eq. (\ref{time_integrated})}, rendering the CME signal to be captured transient in this case. Using a simple impulselike profile for $\mu_5$, we showed that for an axial imbalance source with large enough spatial extension $R$, the nonzero transported charges persists for a timescale comparable to $R$. In case when this characteristic timescale is much larger than the observation time, the magnitude of the charge transport is effectively described by \mbox{Eq. (\ref{CME1})}.

It is also worthwhile to mention that we derived corrections to the homogeneous electric current, which carries structures sensitive to vorticity. Further investigation is needed to decide how this might change the collective behavior of the chiral plasma.

An important lesson we learned is the role of the spatial variation in the axial imbalance, reflected in the gradients of $\ChIm$, which has not been sufficiently addressed in previous works. In the case of homogeneous $\ChIm$ --- which can even be time-dependent --- the UV regularization appears optional since it contributes to the effective action only as a total divergence. Here, we emphasized that in the presence of a nonzero $\nabla\ChIm$, UV regularization is necessary to maintain the local electric charge conservation, i.e., $\partial\cdot J=0$. \REM{Without proper UV regulator, the term causing trouble is $\textstyle \partial\cdot J = \frac{e^2}{2\pi^2}\nabla\ChIm\cdot\ve{B}$.} For any realistic system
there is a boundary where $\ChIm$ changes --- most probably vanishes. If there are EM fields still present around this region, a current
\begin{equation}
\textstyle I=\oint_{\text{boundary}}\mathrm{d}^2\ve{S}\cdot\ve{B}\frac{e^2}{2\pi^2}\,\ChIm 
\end{equation}
is left to be canceled: this is what the UV-term is responsible for in the presented approach. The issue of EM gauge invariance was of course well-recognized in the literature before, see Secs. 2.2--2.4 of Ref. \cite{LandsteinerRev}, for example. The tool to maintain it even in the case of axial anomaly is to add the so-called Bardeen counterterms or Chern-Simons term \cite{Shovkovy3,Shovkovy4}: this is what our fermionic effective action realizes via the PV regulator.

In the weak coupling limit, any time variation of external sources will drive the system out of equilibrium. Therefore, the calculation presented in this work approximates the situation where the characteristic time for the variation of the axial imbalance is shorter than the relaxation time to equilibrium, opposite to the condition assumed in the hydrodynamic regime. There, it is still justified to use \mbox{Eq. (\ref{CME})} -- in the zeroth order of the gradient expansion. As the result of the two orders of limits persists to higher orders in the coupling, the qualitative aspect of our results, say the quenching of the charge transport over a long time may be carried over to the strong coupling regime. 
An important limitation is our simple-minded assumption which models a nonequilibrium axial imbalance by a spacetime dependent axial chemical potential. A more realistic approach without using the notion of chemical potential is to factor in the real time QCD process of the axial charge creation/annihilation inspired by instantons or sphalerons. We hope to be able to report our progress along this line in near future.

\begin{acknowledgments}
D. F. H. and H. C. R. were supported by the Ministry of Science and Technology of China (MSTC) under the ``973'' Project No. 2015CB856904(4) and by the NSFC under Grants No. 11735007 and No. 11875178. M. H. was supported by NSFC under Grant No. 11847242. J. L. is supported by the NSF Grant No. PHY-1913729.
\end{acknowledgments}

\appendix
\begin{widetext}
\section{Anomalous Ward-identity}\label{app:ward}
In this Appendix we show that for homogeneous chiral imbalance the $\wt{\Gamma}^{0ij}_{AVV}$ part of the AVV vertex is completely determined by the axial anomaly. For this we first move $\gamma^5$ into the front in \mbox{Eqs. (\ref{GAVVline1}--\ref{GAVVline6})}. This is done by utilizing the Dirac-structure of the propagators, i.e., $\gamma^5G +G\gamma^5 =g\gamma^5$, where $g$ is a scalar function. Now we can group the terms either as type $GGG$ or $GGg$:
\begin{align}
\left\{\,\,GGG\,\,\right\}&= \nn \\
=\frac{ie^2}{2}\int_p \text{tr}& \left\{ \gamma^5\gamma^i G^C(p+q_2)\gamma^0 G^A(p) \gamma^j G^A(p-q_1) +\right.
\gamma^5\gamma^i G^R(p+q_2)\gamma^0 G^C(p) \gamma^j G^A(p-q_1) +\\
& +\gamma^5\gamma^i G^R(p+q_2)\gamma^0 G^R(p) \gamma^j G^C(p-q_1) +
\gamma^5\gamma^i G^C(p+q_1)\gamma^j G^A(p) \gamma^0 G^A(p-q_2) +\\
& +\gamma^5\gamma^i G^R(p+q_1)\gamma^j G^C(p) \gamma^0 G^A(p-q_2) +
\left. \gamma^5\gamma^i G^R(p+q_1)\gamma^j G^R(p) \gamma^0 G^C(p-q_2) \right\}.
\end{align}
In the next step we set $\ve{q}_2=0$ and utilize \mbox{Eqs. (\ref{WardG1}--\ref{WardG3})} and arrive at:
\begin{align}
\{\,\,GGG\,\,\} =\frac{ie^2}{2}\cdot\frac{1}{q_{20}}
\int_p\text{tr}& \left\{ \gamma^5\gamma^i \left(-G^C(p_0+q_{20},\ve{p})\right) \gamma^j G^A(p-q_1) + \right. \label{anomWardline1}
\gamma^5\gamma^i G^C(p_0,\ve{p}) \gamma^j G^A(p-q_1) +\\ 
& +\gamma^5\gamma^i \left(-G^R(p_0+q_{20},\ve{p})+G^R(p_0,\ve{p})\right) \gamma^j G^C(p-q_1) +
\gamma^5\gamma^i G^C(p+q_1)\gamma^j \left(-G^A(p) +G^A(p_0-q_{20},\ve{p})\right) +\label{anomWardline4}\\
& +\gamma^5\gamma^i G^R(p+q_1)\gamma^j \left(-G^C(p_0-q_{20},\ve{p})\right) +
\left. \gamma^5\gamma^i G^R(p+q_1)\gamma^j G^C(p-q_{20},\ve{p}) \right\}.\label{anomWardline6}
\end{align}
For the vertex function we have the difference of these terms and their UV-limit provided by the PV-terms. This difference is finite, since any dangerous UV behavior is canceled. Therefore one can shift the integration variables in \mbox{Eqs. (\ref{anomWardline1}--\ref{anomWardline6})} and realize that $\left\{\,\,GGG\,\,\right\} -\left\{\,\,GGG\,\,\right\}_\text{PV}\equiv 0$. 

For the rest, we are interested only in the chiral limit. Then since $g_{m=0}=0$, only the PV-terms contribute. The mass scale $M$ in these terms are larger than any other scale in the system. This allows us to replace the fermionic propagators with the noninteracting ones, which also leaves us with $\textstyle g_\text{PV}(p)=\frac{2M}{p^2-M^2}$. One can then make the observation that $q_2$ enters only with $g$, so only in the denominators. Now for the large $M$ limit it is justified to keep $q_1$ only in the nominator, i.e. where it contributes to the spinor structure:
\begin{align}
\wt{\Gamma}^{0ij}_{AVV}(q_1,q_{20},\ve{0}) = -\frac{ie^2}{2}\int_p \text{tr}& \left\{ \gamma^i\gamma^5g^A_\text{PV}(p)\gamma^0G^A_\text{PV}(p)\gamma^j G^A_\text{PV}(p-q_1) +\gamma^i\gamma^5 G^A_\text{PV}(p+q_1)\gamma^j G^A_\text{PV}(p)\gamma^0 g^A_\text{PV}(p) - \left(\,\, A \leftrightarrow R\,\,\right)\right\}(1-2\wt{n}(p_0)).
\end{align}
We can perform the trace and combine the $A$ and $R$ pieces together. Since the only spatial structure is in $\text{tr}\left\{\gamma^i\gamma^5\gamma^0\gamma^j\slashed{q}_1\right\}$, also the directional integration of $\ve{p}$ can be done. The result is the following:
\begin{align}
\wt{\Gamma}^{0ij}_{AVV}(q_1,q_{20},\ve{0}) =& -\frac{1}{16\pi^4}\frac{ie^2}{2}8i\epsilon^{0ijk}q_{1k}2M^2\cdot 4\pi\intlim{p_0}{-\infty}{\infty}\intlim{p}{0}{\infty}p^2\frac{1-2\wt{n}(p_0)}{\left[(p_0-i0^+)^2-p^2-M^2\right]^3} =\\
=& \frac{4e^2}{\pi^3}\epsilon^{0ijk}q_{1k}M^2\intlim{p_0}{-\infty}{\infty}(1-2\wt{n}(p_0))\intlim{p}{0}{\infty}p^2\frac{1}{-8p_0p}\frac{\partial}{\partial p_0}\frac{\partial}{\partial p}\frac{1}{(p_0-i0^+)^2-p^2-M^2} =\\
=& -\frac{e^2}{2\pi^2}\epsilon^{0ijk}q_{1k}M^2\intlim{p_0}{-\infty}{\infty}\frac{\partial}{\partial p_0}\frac{1-2\wt{n}(p_0)}{p_0}\intlim{p}{0}{\infty}i\text{sgn}(p_0)\delta\left(p_0^2-p^2-M^2\right) =\\
=& \frac{e^2}{2\pi^2}i\epsilon^{0ijk}q_{1k}M^2\intlim{p}{0}{\infty}\left(\frac{2\wt{n}'(\sqrt{p^2+M^2})}{p^2+M^2} +\frac{1-2\wt{n}(\sqrt{p^2+M^2})}{(p^2+M^2)^{3/2}}\right) \overset{M\rightarrow\infty}{\longrightarrow} \frac{e^2}{2\pi^2}i\epsilon^{0ijk}q_{1k} +\mathcal{O}(M^{-1}). \label{GAVVhomMu5}
\end{align}
Putting the above expression of \mbox{Eq. (\ref{GAVVhomMu5})} back into \mbox{Eq. (\ref{responseCurrent})} we arrive at the familiar result: the equilibrium CME current of \mbox{Eq. (\ref{CME})} --- with the substitution of $A_{5,0}=-\mu_5$. Although $A_{5,0}$ depends only on time, the magnetic field still can be arbitrary.

Essentially the same argument leads to the anomalous Ward-identity of the vertex function, shown in \mbox{Eq. (\ref{gradJ5})}: for the contraction of $q_{2\rho}\wt{\Gamma}^{\rho\mu\nu}_{AVV}$, one moves the $\gamma^5$ into the front under the trace, then through the same steps as in \mbox{Eqs. (\ref{anomWardline1}--\ref{anomWardline6})} shows that under the integration the regulated version of the expression vanishes --- this time even without the assumption of $\ve{q}_2=0$. So the remaining terms are again the PV ones, leading to $\textstyle q_{2\rho}\frac{e^2}{2\pi^2}i\epsilon^{\rho\mu\nu\sigma}q_{1\sigma}$, i.e., \mbox{Eq. (\ref{gradJ5})}.
\end{widetext}

\section {Relation to the Maxwell-Chern-Simons Electrodynamics}\label{app:MCSeldin}
Let us briefly return to the phenomenology of the QCD matter. 
We already mentioned that the local $CP-$violation is encoded in $A_5$. According to the anomalous Ward-identity, this contributes to $\partial\cdot J_5$ by a source term $\textstyle\frac{e^2}{6\pi^2}\ve{E}_5\cdot\ve{B}_5$. We also know, that the gauge sector of QCD has its contribution to the balance equation as $\textstyle\frac{N_fg^2}{8\pi^2}\ve{E}_a\cdot\ve{B}^a$. Now we assume that the gluon sector affects the EM transport through the axial imbalance, but there is no backreaction. After the dynamics of the gauge fields is integrated out an effective action like in \mbox{Eq. (\ref{EMLagrange})} should emerge. Although we do not know how the quark-gluon vertices map into the axial gauge fields, we assume the matching of the previously mentioned sources for $\partial\cdot J_5$. 

Setting the fully dynamical origin of $A_5$ aside, what we know that it originates from the vacuum sectors with nontrivial topology (which can be inhomogeneously distributed in space). A minimal approach to model this is to add a so-called axion term to the original EM Lagrangian:
\begin{align}
\mathcal{L}_{\theta} =& -\frac{1}{4}F_{\mu\nu}F^{\mu\nu} + \overline{\psi}\gamma^\mu\left(i\partial_\mu -eA_\mu \right)\psi +\frac{e^2}{16\pi^2}\epsilon^{\mu\nu\alpha\beta}\theta F_{\mu\nu}F_{\alpha\beta}.
\end{align}
At this point we can \textit{impose the anomalous Ward-identity} to the system and identify the current in terms of the fermionic fields:

\begin{align}
\frac{e^2}{16\pi^2}\epsilon^{\mu\nu\alpha\beta} F_{\mu\nu}F_{\alpha\beta} \overset{!}{=} \partial_\mu J_5^\mu = \partial_\mu\overline{\psi}\gamma^\mu\gamma^5\psi, \label{anomWIQED}
\end{align}
which after partial integration leads to $\mathcal{L}$, with the axial vector potential $A_{5,\mu}=\partial_\mu\theta$. This special form of $A_5$ renders the electric response to a simple form, solely determined by the anomaly. Using \mbox{Eq. (\ref{responseCurrent})} and \mbox{Eq. (\ref{gradJ5})}, we arrive at the following expressions:

\begin{align}
J^0 =& \frac{e^2}{2\pi^2}\nabla\theta\cdot\ve{B}, \label{MCScurrent1}\\
\ve{J} =& \frac{e^2}{2\pi^2}\dot{\theta}\ve{B}, \label{MCScurrent2}
\end{align} 
which are the well-known equations of motion of the Maxwell-Chern-Simons electrodynamics \cite{CSeldinExample} in the special case of constant $\ve{B}$ and vanishing $\ve{E}$. \REM{The above result is in agreement with the analysis of the vertex function which has led to \mbox{Eq. (\ref{CME1})}, as $\dot{\theta}=A_{5,0}$. As we pointed out earlier, the vector current expression differs by its sign from \mbox{Eq. (\ref{CME})} if one identifies the temporal component of the axial field by $-\mu_5$. One should, however, keep in mind that it is not required for the system to be in thermal equilibrium.} This simple form of the electric current is the consequence of the anomalous Ward-identity at the level of the vertex function, i.e. \mbox{Eq. (\ref{gradJ5})}, therefore it is 
not sensitive to the details of the underlying fermionic dynamics in this case. Although this statement remains true even if dynamical EM fields are present, the axial current is not tied to the anomaly anymore. As can be seen from \mbox{Eq. (\ref{J5linres})}, the first and third terms vanish if $A_5$ is a pure gradient, the second term is sensitive to the EM-fields only. Therefore the IR behavior of the AVV vertex becomes important for $J_5$. But since $F_5^{\mu\nu}=\partial^\mu A_5^\nu -\partial^\nu A_5^\mu\equiv 0$ for $A_5^\mu=\partial^\mu\theta$, there is no chiral charge generation. So if $\ve{E}\cdot\ve{B}=0$ and initially $Q_5$ is zero, there is still a CME-like current. This might seem troubling, however, one quickly realizes that $\mathcal{L}_{\theta}$ is actually not the system we are interested in. QCD has a $\theta-$term for the \textit{gluonic} sector. The effective action for the EM sector, only indicating the $G\wt{G}$ part of the gluon field strength, looks like this:

\begin{align}
\mathcal{L}_{\theta,\text{QCD}} =& -\frac{1}{4}F_{\mu\nu}F^{\mu\nu} + \overline{\psi}\gamma^\mu\left(i\partial_\mu -eA_\mu \right)\psi +\frac{g^2}{32\pi^2}\theta G_{\mu\nu}^a\wt{G}^{\mu\nu}_a.
\end{align}
Now, \mbox{Eq. (\ref{anomWIQED})} is not the right anomaly relation for QCD. Instead, one has

\begin{align}
\frac{e^2}{16\pi^2}\epsilon^{\mu\nu\alpha\beta} F_{\mu\nu}F_{\alpha\beta} +\frac{e^2}{32\pi^2}\epsilon^{\mu\nu\alpha\beta} G_{a,\mu\nu}G_{a,\alpha\beta} \overset{!}{=} \partial_\mu J_5^\mu = \partial_\mu\overline{\psi}\gamma^\mu\gamma^5\psi, \label{anomWIQCD}
\end{align}
which leaves us not only with $\mathcal{L}$, rather with $\textstyle \mathcal{L} - \theta \frac{e^2}{16\pi^2}\epsilon^{\mu\nu\alpha\beta} F_{\mu\nu}F_{\alpha\beta}$. We can still recast the remaining term as $\textstyle \frac{e^2}{2\pi^2}\frac{\theta}{4}\partial_\mu J^\mu_{CS}$, with the so-called Chern-Simons current $J^\mu_{CS}=\epsilon^{\mu\nu\rho\sigma}A_\nu F_{\rho\sigma}$. It is straightforward to show, that the same components of $J$ generated from $J_{CS}$ as in \mbox{Eqs. (\ref{MCScurrent1},\ref{MCScurrent2})}, but with opposite sign. That is, the CME-like current vanishes. Our intuition therefore restored, there is no CME with zero $Q_5$.

What we can conclude is that the simplest way of taking topological effects into account, namely by adding the $\theta$-term, is not sufficient. The reason is that a pure gradient axial gauge field does not contribute to the chiral charge balance equation. Nevertheless, a possible $\theta-$term still can cause fluctuations both in the vector and the axial currents.
\begin{widetext}

\section{Trace calculation}\label{app:trace}
In this Appendix we give the details of calculating the trace $\textstyle\text{tr}\left\{\gamma^5\gamma^iG(p+q)\gamma^0\gamma^5G(p)\gamma^j\frac{\partial}{\partial p_k}G(p)\right\}\epsilon^{ljk}B_l$ for noninteracting fermions.
Using the explicit form of the propagator, the trace expression can be written like this:
\begin{align}
@=\text{tr}\left\{\gamma^i(\slashed{p}+\slashed{q}+m)\gamma^0\gamma^5(\slashed{p}+m)\gamma^j \frac{\partial}{\partial p_k}\frac{\slashed{p}+m}{p^2-m^2} \right\}\epsilon^{ljk}B_l= & \frac{1}{p^2-m^2}\underbrace{\epsilon^{ljk}B_l\text{tr}\left\{\gamma^i(\slashed{p}+\slashed{q}+m)\gamma^0\gamma^5(\slashed{p}+m)\frac{\gamma^j\gamma^k-\gamma^k\gamma^j}{2}\right\}}_{=:I} +\\
+ & \frac{2}{(p^2-m^2)^2}\underbrace{\text{tr}\left\{\gamma^i(\slashed{p}+\slashed{q}+m)\gamma^0\gamma^5(\slashed{p}+m)\gamma^j (\ve{p}\times\ve{B})^j (\slashed{p}+m)\right\}}_{=:II}.
\end{align}
The detailed evaluation of term $I$:
\begin{align}
I =& 2i\text{tr}\left\{\gamma^i(\slashed{p}+\slashed{q}+m)\gamma^0\gamma^5(\slashed{p}+m)\gamma^0\gamma^l\gamma^5\right\}B^l = 2i \text{tr}\left\{\gamma^i(\slashed{p}+\slashed{q}+m)\gamma^0(-\slashed{p}+m)\gamma^0\gamma^l\right\}B^l =\\
 =& 2i m^2 B^l\text{tr}\left\{\gamma^i\gamma^0\gamma^0\gamma^l\right\} -2iB^l\left(p^0 \text{tr}\left\{\gamma^i(\slashed{p}+\slashed{q})\gamma^0\gamma^l\right\} +p^m\text{tr}\left\{	\gamma^i(\slashed{p}+\slashed{q})\gamma^m\gamma^l\right\}\right) =\\
 =& 8i\left[ m^2\eta^{il}B^l -B^l\left(p^0(p+q)_\alpha\left[\eta^{i\alpha}\eta^{0l}-\eta^{i0}\eta^{\alpha l} +\eta^{il}\eta^{\alpha 0}\right] -p^m(-4)(p+q)_\alpha\left[\eta^{i\alpha}\eta^{ml}-\eta^{im}\eta^{\alpha l}+\eta^{il}\eta^{\alpha m}\right]\right) \right]= \\
 =& -8i\left( B^i\left[m^2-p^0(p^0+q^0)-\ve{p}\cdot(\ve{p}+\ve{q})\right]-q^i\ve{B}\cdot\ve{p} +p^i\ve{B}\cdot\ve{q}\right).
\end{align}
Evaluating term $II$:
\begin{align}
II =& (\ve{p}\times\ve{B})^j\text{tr}\left\{\gamma^i(\slashed{p}+\slashed{q}+m)\gamma^0\gamma^5(\slashed{p}+m)\gamma^j(\slashed{p}+m)\right\} =\\
 =& (\ve{p}\times\ve{B})^j\text{tr}\left\{\gamma^i(\slashed{p}+\slashed{q}+m)\gamma^0\gamma^5(\slashed{p}+m)(2p^j -(\slashed{p}-m))\gamma^j\right\} = \\
 =& -(p^2-m^2)\text{tr}\left\{\gamma^i(\slashed{p}+\slashed{q}+m)\gamma^0\gamma^5\gamma^j\right\}(\ve{p}\times\ve{B})^j\\
 =& 4i(p^2-m^2)\left[(\ve{p}+\ve{q})\times\ve{p}\times\ve{B}\right]^i =4i(p^2-m^2)\left[p^i\ve{B}\cdot(\ve{p}+\ve{q}) -B^i(\ve{p}+\ve{q})\cdot\ve{p}\right].
\end{align}
Finally, putting the contributions together we get:
\begin{align}
@ =& \frac{-8i}{p^2-m^2}\left[B^i\left(m^2-p_0(p_0+q_{0})\right)-(p^i+q^i)\ve{B}\cdot\ve{p}\right].
\end{align}

\section{Constant $\ve{B}$ calculation}\label{app:constB}
Here we give the detailed calculations leading to the conductivity $\overline{\sigma}^i_A(t,\ve{q})$ in \mbox{Eq. (\ref{constBrespFunc})}.

\subsection{Manipulations of $\frac{\partial\mathfrak{g}^{ij}_m}{\partial q_{1k}}$}
First the trace has been calculated. Then we regrouped the terms of \mbox{Eqs. (\ref{gAVVconstBline1}--\ref{gAVVconstBline3})} in a combinations of $AAA\,+\,RAA$ and $RRR\,+\,RAA$. After recasting the higher order pole-contributions as derivatives and also changing integration variables so one can separate $\textstyle (1-2\wt{n}(p_0))\frac{\text{sgn}(p_0)}{2|\ve{p}|}$, the resulting expression is
\begin{align}
\frac{\partial\mathfrak{g}^{ij}_m}{\partial q_{1k}}\epsilon^{ljk}B_l =& -16\pi e^2\int_p(1-2\wt{n}(p_0))\text{sgn}(p_0)\frac{1}{2|\ve{p}|}\times \nonumber \\
&\times \left[ \left(B^i\left(m^2-p_0(p_0+q_0)\right)-p^i\ve{B}\cdot\ve{p}\right)\frac{\partial}{\partial|\ve{p}|}\frac{\delta\left(p_0^2-\ve{p}^2-m^2\right)}{(p_0+q_0+i0^+)^2-(\ve{p}+\ve{q})^2-m^2} +\right. \\
& -\left\{p^i\ve{B}\cdot\ve{q}\frac{\partial}{\partial|\ve{p}|}\left(\frac{1}{(p_0+q_0+i0^+)^2-(\ve{p}+\ve{q})^2-m^2}\right) +\frac{q^i\ve{B}\cdot\ve{p}}{(p_0+q_0+i0^+)^2-(\ve{p}+\ve{q})^2-m^2} \frac{\partial}{\partial |\ve{p}|}\right\}\delta\left(p_0^2-\ve{p}^2-m^2\right) + \\
&\left. -\frac{\frac{\ve{p}\cdot\ve{q}}{p^2}}{1+\frac{\ve{p}\cdot\ve{q}}{p^2}}\left(B^i\left(m^2-p_0(p_0+q_0)\right) -p^i\ve{B}\cdot(\ve{p}+\ve{q})\right)\frac{\partial}{\partial |\ve{p}|}\left(\frac{1}{(p_0+q_0+i0^+)^2-(\ve{p}+\ve{q})^2-m^2}\right)\delta\left(p_0^2-\ve{p}^2-m^2\right)\right].
\end{align}
\mbox{Equation (\ref{gAVVconstBderivation1})} is the direct consequence of the above. We proceed by simplifying the angular integration by separating $\ve{B}$ into components parallel and perpendicular to $\ve{q}$:
\begin{align}
\frac{\partial\mathfrak{g}^{ij}_m}{\partial q_{1k}}\epsilon^{ljk}B_l\overset{\ve{B}=B_\parallel\frac{\ve{q}}{q}+\ve{B}_\perp}{=} & \frac{e^2}{\pi^2}B_\parallel \widehat{q}^i\intlim{p_0}{-\infty}{\infty}\text{sgn}(p_0)(1-2\wt{n}(p_0))\intlim{p}{0}{\infty}\intlim{x}{-1}{1} \times \nonumber \\
& \times \left[ \frac{\partial}{\partial p}\left( p\left(\left(m^2-p_0(p_0+q_0)\right)-p^2x^2\right)- qxp^2\right) \frac{\delta\left(p_0^2-p^2-m^2\right)}{(p_0+q_0+i0^+)^2-p^2-q^2-2pqx-m^2} + \right. \\
& \left. +2qpx\left(m^2-p_0(p_0+q_0)-qxp -p^2x^2\right)\frac{\delta\left(p_0^2-p^2-m^2\right)}{\left[(p_0+q_0+i0^+)^2-p^2-q^2-2qpx-m^2\right]^2} \right]+ \\
& +\frac{e^2}{\pi^2}B_\perp^i \intlim{p_0}{-\infty}{\infty}\text{sgn}(p_0)(1-2\wt{n}(p_0))\intlim{p}{0}{\infty}\intlim{x}{-1}{1}\times\nonumber \\
\times & \left[ \frac{\frac{\partial}{\partial p}\left(p\left(m^2-p_0(p_0+q_0))-\frac{1-x^2}{2}p^2\right)\right)}{(p_0+q_0+i0^+)^2-p^2-q^2-2qpx -m^2} +2qpx\frac{m^2-p_0(p_0+q_0)-\frac{1-x^2}{2}p^2}{\left[(p_0+q_0+i0^+)^2-p^2-q^2-2qpx -m^2\right]^2}\right]\delta\left(p_0^2-p^2-m^2\right).
\end{align}

Regrouping terms leads to:
\begin{align}
\frac{\partial\mathfrak{g}^{ij}_m}{\partial q_{1k}}\epsilon^{ljk}B_l= & -\frac{e^2}{\pi^2}\intlim{p_0}{-\infty}{\infty}\text{sgn}(p_0)(1-2\wt{n}(p_0))\intlim{p}{0}{\infty}\delta\left(p_0^2-p^2-m^2\right) \intlim{x}{-1}{1}\times \nonumber \\
& \times \left\{ B_\parallel^i \left[ \frac{p_0q_0 +2qpx +(3x^2+1)p^2}{(p_0+q_0+i0^+)^2-p^2-q^2-2qpx-m^2} + \frac{2qpx\left(p_0q_0+qpx+(x^2+1)p^2\right)}{\left[(p_0+q_0+i0^+)^2-p^2-q^2-2qpx-m^2\right]^2} \right] +\right. \\
& \left. + B_\perp^i \left[ \frac{p_0q_0 +\left[\frac{3}{2}(1-x^2)+1\right]p^2}{(p_0+q_0+i0^+)^2-p^2-q^2-2qpx-m^2} + \frac{2qpx\left(p_0q_0+\left[\frac{1-x^2}{2}+1\right]p^2\right)}{\left[(p_0+q_0+i0^+)^2-p^2-q^2-2qpx-m^2\right]^2} \right] \right\},
\end{align}
which gives us \mbox{Eq. (\ref{constBrespMomspaveLine2})} after part of the expression is written as a total derivative with respect to $x$.

\subsection{Derivation of \mbox{Eq. (\ref{chiral_term})}}

The Fourier transformation of \mbox{Eq. (\ref{constBrespMomspaveLine2})} with respect to $q_0$ can be calculated readily: 

\begin{align}
\intlimB{q_{0}}{-\infty}{\infty}e^{iq_{0}t} \frac{\partial\mathfrak{g}^{ij}}{\partial q_{1k}}\epsilon^{ljk}B_l =& \theta(-t)\frac{e^2}{\pi^2}  \intlim{p_0}{-\infty}{\infty}(1-2\wt{n}(p_0))e^{-ip_0t}\intlim{p}{0}{\infty}\text{sgn}(p_0)\delta\left(p_0^2-p^2-m^2\right)\intlim{x}{-1}{1}\times \nonumber \\
\times & \left\{ \frac{\partial}{\partial x}x \left[ B_\parallel^i \left(\frac{p_0^2\sin\left(t\sqrt{p^2+q^2+2qpx+m^2}\right)}{\sqrt{p^2+q^2+2qpx+m^2}} +ip_0\cos\left(t\sqrt{p^2+q^2+2qpx+m^2}\right) + \right.\right.\right. \\
& \left. -\frac{((1+x^2)p^2+qpx)\sin\left(t\sqrt{p^2+q^2+2qpx+m^2}\right)}{\sqrt{p^2+q^2+2qpx+m^2}} \right) +\\
& +B_\perp^i\left( \frac{p_0^2\sin\left(t\sqrt{p^2+q^2+2qpx+m^2}\right)}{\sqrt{p^2+q^2+2qpx+m^2}} +ip_0\cos\left(t\sqrt{p^2+q^2+2qpx+m^2}\right) +\right. \\
& \left. \left. \left. -\left(\frac{1-x^2}{2}+1\right)p^2\frac{\sin\left(t\sqrt{p^2+q^2+2qpx+m^2}\right)}{\sqrt{p^2+q^2+2qpx+m^2}}\right)\right] - B_\perp^i \frac{p^2\sin\left(t\sqrt{p^2+q^2+2qpx+m^2}\right)}{\sqrt{p^2+q^2+2qpx+m^2}}\right\}.
\end{align}
After the $p_0-$ and $x-$integrations we get:

\begin{align}
\#=&\intlimB{q_0}{-\infty}{\infty}e^{iq_0t}\frac{\partial\mathfrak{g}^{ij}_{m=0}}{\partial q_{1k}}\epsilon^{ljk}B_l =\theta(-t)\frac{e^2}{\pi^2}\intlim{p}{0}{\infty}\frac{1-2\wt{n}(\Omega)}{2\Omega}\times \nonumber\\
& \times\left\{ B_\parallel^i\left( 2\cos(\Omega t)\left[ \frac{(\Omega^2-2p^2-qp)\sin\left(t\sqrt{(p+q)^2+m^2}\right)}{\sqrt{(p+q)^2+m^2}} + \frac{(\Omega^2-2p^2+qp)\sin\left( t\sqrt{(p-q)^2+m^2}\right)}{\sqrt{(p-q)^2+m^2}}\right] +\right. \right. \\
& \left. +2\Omega\sin(\Omega t)\left[\cos\left(t\sqrt{(p+q)^2+m^2}\right) +\cos\left(t\sqrt{(p-q)^2+m^2}\right) \right]\right) +\\
& +B_\perp^i \left( 2m^2\cos(\Omega t)\left[\frac{\sin\left(t \sqrt{(p+q)^2+m^2}\right)}{\sqrt{(p+q)^2+m^2}} +\frac{\sin\left(t \sqrt{(p-q)^2+m^2}\right)}{\sqrt{(p-q)^2+m^2}} \right] +\right. \\
&\left. \left. +\left(2\Omega\sin(\Omega t)+\frac{2p\cos(\Omega t)}{qt}\right)\cos\left(t\sqrt{(p+q)^2+m^2}\right) + \left(2\Omega\sin(\Omega t)-\frac{2p\cos(\Omega t)}{q t}\right)\cos\left(t\sqrt{(p-q)^2+m^2}\right) \right)\right\},
\end{align}
where $\Omega=\sqrt{p^2+m^2}$. Then we take the chiral limit and collect terms carefully. Only a $\ve{B}_\perp$ contribution remains:

\begin{align}
\#= & \theta(-t)\frac{e^2}{\pi^2}\intlim{p}{0}{\infty}(1-2\wt{n}(p)) \left\{ B_\parallel^i\left[-\cos(pt)\underbrace{\left(\sin(t(p+q)) +\sin(t(p-q))\right)}_{=2\cos(qt)\sin(pt)} +\sin(pt)\underbrace{\left(\cos(t(p+q))+\cos(t(p-q))\right)}_{=2\cos(qt)\cos(pt)}\right] +\right. \\
& \left. +B_\perp^i \left[ \left(\sin(pt)+\frac{\cos(pt)}{qt}\right)\cos(t(p+q)) + \left(\sin(pt)-\frac{\cos(pt)}{qt}\right)\cos(t(p-q))\right]\right\}= \\
=& \theta(-t)\frac{e^2}{\pi^2}B_\perp^i\intlim{p}{0}{\infty}(1-2\wt{n}(p)) \left[ \sin(pt)2\cos(pt)\cos(q t) +\frac{1}{qt}\cos(pt)(-2)\sin(pt)\sin(qt)\right] = \\
=& \theta(-t)\frac{e^2}{\pi^2} (B^i-\widehat{q}^i(\ve{B}\cdot\widehat{\ve{q}})) t\frac{\partial}{\partial t}\left(\frac{\sin(qt)}{qt}\right)T\intlim{y}{0}{\infty}(1-2n_{FD}(y))\sin(2yTt),
\end{align}
resulting in \mbox{Eq. (\ref{chiral_term})} in the end.

\subsection{Derivation of \mbox{Eq. (\ref{PV_term})}}

Scaling the loop momentum $p$ of \mbox{Eq. (\ref{constBrespMomspaveLine2})} by $p=My$ and taking the limit $\to 0$, we find that:

\begin{align}
\left.\frac{\partial \mathfrak{g}^{ij}_{M\rightarrow\infty}}{\partial q_{1k}}\right|_{q_1=0}\epsilon^{ljk}B_l
\approx & -\frac{e^2}{\pi^2}\intlim{y_0}{-\infty}{\infty}\intlim{y}{0}{\infty}\frac{\delta(y_0-\sqrt{y^2+1})+\delta(y_0+\sqrt{y^2+1})}{2\sqrt{y^2+1}}\frac{1}{2}\intlim{x}{-1}{1}\nonumber \\
& \left\{ \frac{\partial}{\partial x}x\left[ (B_\parallel^i+B_\perp^i)\frac{q_0}{q_0-\frac{yqx}{y_0}+i0^+} +B_\parallel^i \frac{qyx}{y_0}\frac{1}{q_0-\frac{yqx}{y_0}+i0^+} -\left(B_\parallel^i(x^2+1)+B_\perp^i\right)y^2\frac{q_0^2-q^2}{2y_0^2}\frac{y_0}{yx}\frac{\partial}{\partial q}\frac{1}{q_0-\frac{yqx}{y_0}+i0^+} \right] + \right. \\
&\left. -B_\perp^i y^2\frac{q_0^2-q^2}{2y_0^2} \frac{y_0}{yq}\frac{\partial}{\partial x}\frac{1}{q_0-\frac{yqx}{y_0}+i0^+}\right\} =: @
\end{align}
Then comes the Fourier transform and the leftover integration. Step-by-step it is done as follows:

\begin{align}
\intlimB{q_0}{-\infty}{\infty}e^{iq_0t}@ =& \intlim{x}{-1}{1}\left\{ \frac{\partial}{\partial x}x\left[ (B_\parallel^i+B_\perp^i)\frac{\partial}{\partial t}\left(-\theta(-t)e^{i\frac{xqy}{y_0}t}\right) -B_\parallel^i \frac{xyq}{y_0}i\theta(-t)e^{i\frac{xqy}{y_0}t} -\left(\frac{B_\parallel^i}{2}\frac{x^2+1}{x}+\frac{B_\perp^i}{2} \frac{1}{x}\right)\frac{y}{y_0}\left(\frac{\partial^2}{\partial t^2}+q^2\right)\left(-\theta(-t)\frac{xyt}{y_0}e^{i\frac{xqy}{y_0}t}\right) \right] +\right.\\
& \left. +\frac{B_\perp^i}{2}\frac{y}{y_0}\frac{1}{q}\frac{\partial}{\partial x}\left(\frac{\partial^2}{\partial t^2}+q^2\right)\left(-i\theta(-t)e^{i\frac{xqy}{y_0}t}\right) \right\} =:\intlim{x}{-1}{1}\frac{\partial}{\partial x}[\dots]. \label{FTgMconstB}
\end{align}	
At this point one has to account for the $t$-derivatives in the integrand:

\begin{align}
[\dots] 
=& -B_\parallel^i \left\{-\left(1-\frac{x^2+1}{2}\frac{y^2}{y_0^2}\right)\delta(t)x +\theta(-t)\left[ \left(1-\frac{x^2+1}{2}\frac{y^2}{y_0^2}\right)2iq\frac{y}{y_0}x^2-\left(1-x^2\frac{y^2}{y_0^2}\right)tq^2\frac{y^2}{y_0^2}\frac{x^2+1}{2}x\right]e^{iqt\frac{y}{y_0}x} \right\}+ \\
& -B_\perp^i\left\{-\left(1-\frac{y^2}{y_0^2}\right)\delta(t)x+\theta(-t)\left[\left(1-\frac{y^2}{y_0^2}\right)iq\frac{y}{y_0}x^2 +\left(1-x^2\frac{y^2}{y_0^2}\right)\left(-q^2t\frac{y^2}{y_0^2}\frac{x}{2} +iq\frac{y}{y_0}\frac{1}{2}\right)\right]e^{iqt\frac{y}{y_0}x} \right\},
\end{align}
which we put back into the \mbox{Eq. (\ref{FTgMconstB})} to gain the expression

\begin{align}
\intlimB{q_0}{-\infty}{\infty}e^{iq_0t}@ =& B_\parallel^i \left(1-\frac{y^2}{y_0^2}\right)\left(2\delta(t) -\theta(-t)\left[-4q\frac{y}{y_0}\sin\left(\frac{y}{y_0}qt\right) -2q^2t\frac{y^2}{y_0^2}\cos\left(\frac{y}{y_0}qt\right)\right]\right) +\\
& +B_\perp^i\left(1-\frac{y^2}{y_0^2}\right)\left( 2\delta(t) -\theta(-t)\left[-3q\frac{y}{y_0}\sin\left(\frac{y}{y_0}qt\right) -q^2t\frac{y^2}{y_0^2}\cos\left(\frac{y}{y_0}qt\right)\right]\right) =\\
=& \left(1-\frac{y^2}{y_0^2}\right)\left(2(B_\parallel^i+B_\perp^i)\delta(t)-\theta(-t)\left[(2B_\parallel^i+B_\perp^i)\frac{\partial^2}{\partial t^2}\left\{t\cos\left(\frac{y}{y_0}qt\right)\right\} +B_\perp^i \frac{\partial}{\partial t}\cos\left(\frac{y}{y_0}qt\right)\right]\right).
\end{align}
After collecting terms, \mbox{Eq. (\ref{PV_term})} follows.

\section{Response functions in coordinate space}\label{app:coordResp}

It is possible to perform the inverse Fourier transform of \mbox{Eq. (\ref{constBrespFunc})} from the spatial momentum $\ve{q}$ to position $\ve{r}$ as well. One needs to carefully treat term-by-term the different $\ve{q}$-contributions:

\begin{align}
\ve{B}q\sin qt, & \,\,\,\,\widehat{\ve{q}}(\ve{B}\cdot\widehat{\ve{q}})\sin qt, \nonumber\\
\ve{B}\frac{\partial}{\partial t}\left(\frac{\sin qt}{qt}\right), & \,\,\,\,\widehat{\ve{q}}(\ve{B}\cdot\widehat{\ve{q}})\frac{\partial}{\partial t}\left(\frac{\sin qt}{qt}\right). \nonumber
\end{align}
After tedious but straightforward calculation, the following expression emerges:

\begin{align}
\ve{J}(t,\ve{r}) =& \actSign\intlim{^2\widehat{\ve{r'}}}{}{}\intlim{r'}{0}{\infty}\frac{e^2}{2\pi^2}\left\{ \ve{B}\ChIm(t,\ve{r'}+\ve{r})\frac{r'}{4\pi}\left(-\frac{\partial}{\partial r'}\delta(r')\right) +\right.\\
& +\frac{\ve{B}}{2}\frac{(r')^2}{4\pi r'}\left[ \ChIm(t,\ve{r}+\ve{r'})\delta'(r') +\partial_1^2\ChIm(t-r',\ve{r}+\ve{r'})+\frac{\partial_1(\ChIm(t-r',\ve{r}+\ve{r'})\wt{f}(-r'))}{r'}\right] + \\
& +\frac{\ve{B}}{2}\frac{1}{4\pi}\left[ \delta(r)\ChIm(t,\ve{r}+\ve{r'})- \partial_1\ChIm(t-r',\ve{r}+\ve{r'})-\frac{\ChIm(t-r',\ve{r}+\ve{r'})-\ChIm(t,\ve{r}+\ve{r'})}{r'}\right] +\\
& +\frac{\widehat{\ve{r}}'(\ve{B}\cdot\widehat{\ve{r}}')}{2}\frac{1}{4\pi}\left[r'\delta'(r')\ChIm(t,\ve{r}+\ve{r'})+r'\partial_1^2\ChIm(t-r',\ve{r}+\ve{r'}) -3\delta(r')\ChIm(t,\ve{r}+\ve{r'})+3\partial_1\ChIm(t-r',\ve{r}+\ve{r'}) \right.+\nonumber \\
&\left. \left. + 3\frac{\ChIm(t-r',\ve{r}+\ve{r'})-\ChIm(t,\ve{r}+\ve{r'})}{r'} -\partial_1(\ChIm(t-r',\ve{r}+\ve{r'})\wt{f}(-r'))\right] \right\},
\end{align}
where $\wt{f}(x)=f(xT)$.
This can be cast in a more compact form by collecting terms into the following groups:

\begin{align}
\ve{J}(t,\ve{r}) =& \actSign\frac{e^2}{2\pi^2}\left\{\ve{B}\ChIm(t,\ve{r}) -\frac{2}{3}\ve{B}\ChIm(t,\ve{r}) + \specLB\right. \\
& +\frac{1}{8\pi}\intlim{^2\widehat{\ve{r'}}}{}{}\intlim{r'}{0}{\infty}\left[\left(r'\partial_1^2\ChIm(t-r',\ve{r}+\ve{r'})\right)(\ve{B}+\widehat{\ve{r}}'(\ve{B}\cdot\widehat{\ve{r}}')) + \specLB\right.\\
& -\left(\partial_1\ChIm(t-r',\ve{r}+\ve{r'}) +\frac{\ChIm(t-r',\ve{r}+\ve{r'})-\ChIm(t,\ve{r}+\ve{r'})}{r'}\right)(\ve{B}-3\widehat{\ve{r}}'(\ve{B}\cdot\widehat{\ve{r}}')) + \\
& +\partial_1\left(\ChIm(t-r',\ve{r}+\ve{r'})\wt{f}(-r')\right)(\ve{B}-\widehat{\ve{r}}'(\ve{B}\cdot\widehat{\ve{r}}')) \left]\left\}. \specLB\right.\right. \label{constBrespCoordSpace}
\end{align}

\subsection{CME response to pointlike perturbation}
The formula given in \mbox{Eq. (\ref{constBrespCoordSpace})} can be better understood via an example. For that, let us suppose $\ChIm$ is well-localized in space as $\ChIm(t,\ve{r})=\ChIm(t)\delta^{(3)}(\ve{r})$. All the remaining integration can be done with the aid of the delta-function. The result is the following expression:

\begin{align}
\ve{J}(t,\ve{r}) = \actSign\left.\frac{e^2}{2\pi^2}\right\{ & \frac{1}{3}\ve{B}\ChIm(t)\delta^{(3)}(\ve{r}) +\nonumber \\
& +\frac{1}{2}\left[\frac{\ChIm''(t-r)}{r}(\ve{B}+(\ve{B}\cdot\widehat{\ve{r}})\widehat{\ve{r}}) +\right.\nonumber \\
& -\left(\frac{\ChIm'(t-r)}{r^2}+\frac{\ChIm(t-r)-\ChIm(t)}{r^3}\right)(\ve{B}-3(\ve{B}\cdot\widehat{\ve{r}})\widehat{\ve{r}}) +\nonumber \\
&\left.\left. + \frac{\ChIm'(t-r)f(rT)-\ChIm(t-r)Tf'(rT)}{r^2}(\ve{B}-(\ve{B}\cdot\widehat{\ve{r}})\widehat{\ve{r}})\right]\right\}. \label{constBrespCoordSpacePointLikeMu}
\end{align}
There is a contribution centered at the origin, the one third of the CME current in \mbox{Eq. (\ref{CME1})}. The other contributions carry various position dependence, depicted in Fig. \ref{fig:vecFields}. It is interesting to observe a stationary source, i.e., $\ChIm(t)\equiv\ChIm$. In this case the only contribution except the delta-term is a finite temperature one: the last term in the 4th line of \mbox{Eq. (\ref{constBrespCoordSpacePointLikeMu})}. Averaging the current over a small region around the origin we get back the expression we already derived previously in \mbox{Eq. (\ref{constBrespQuenchedLTAvr})}: the long-time limit of the averaged current after a quench.

\begin{figure}[!h]
	\begin{tabular}{m{0.3\linewidth}m{0.3\linewidth}m{0.3\linewidth}}
		\includegraphics[width=0.85\linewidth]{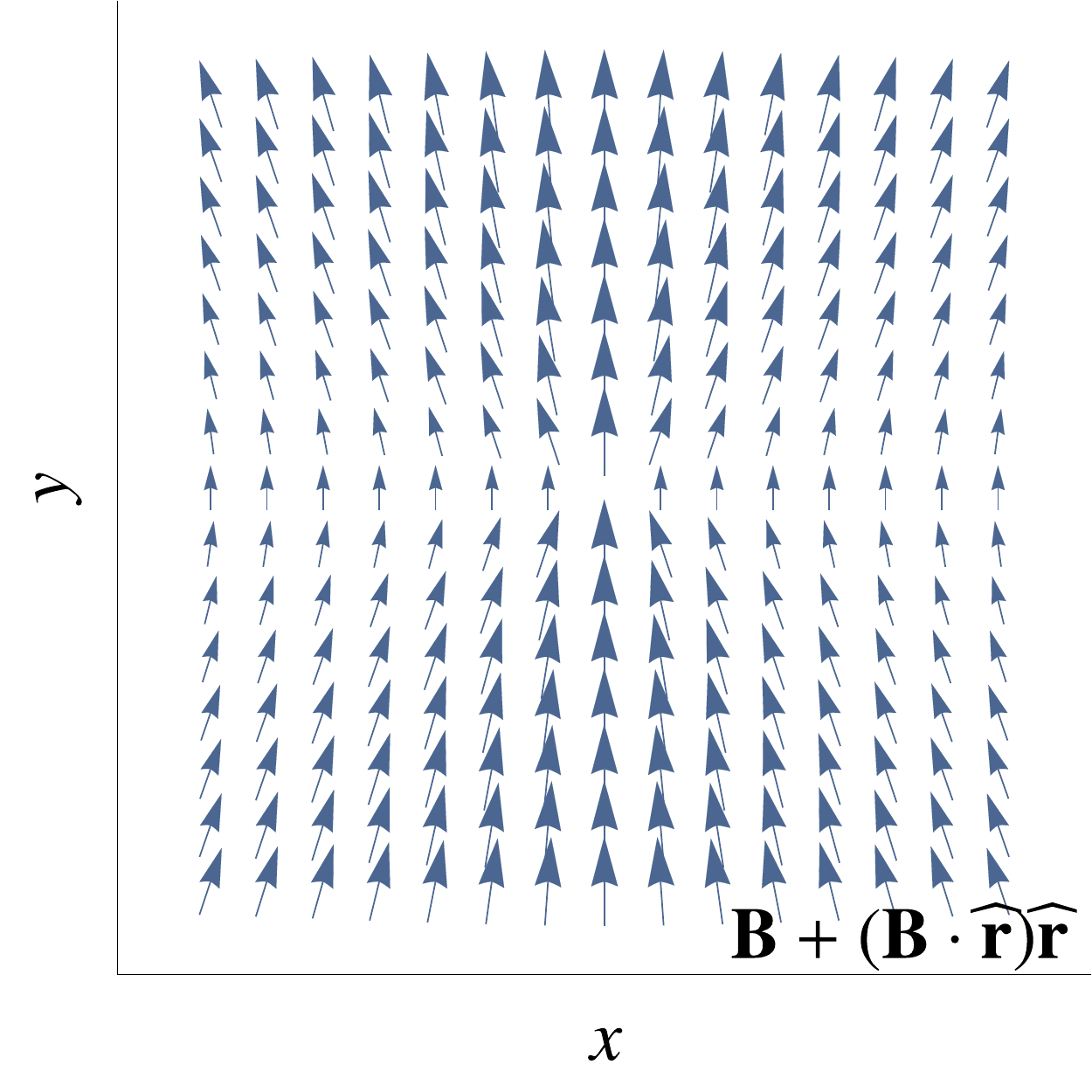} &
		\includegraphics[width=0.85\linewidth]{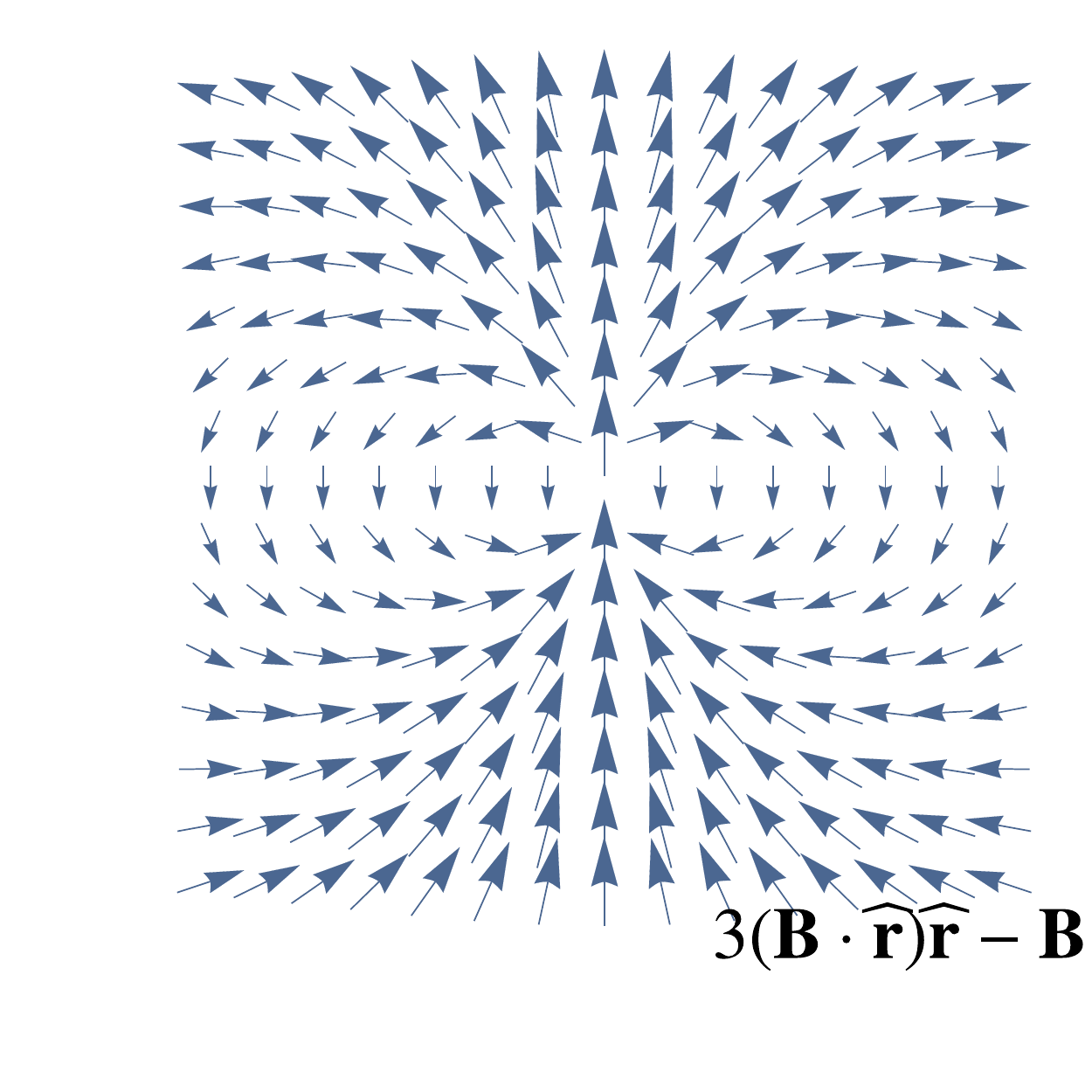} &
		\includegraphics[width=0.85\linewidth]{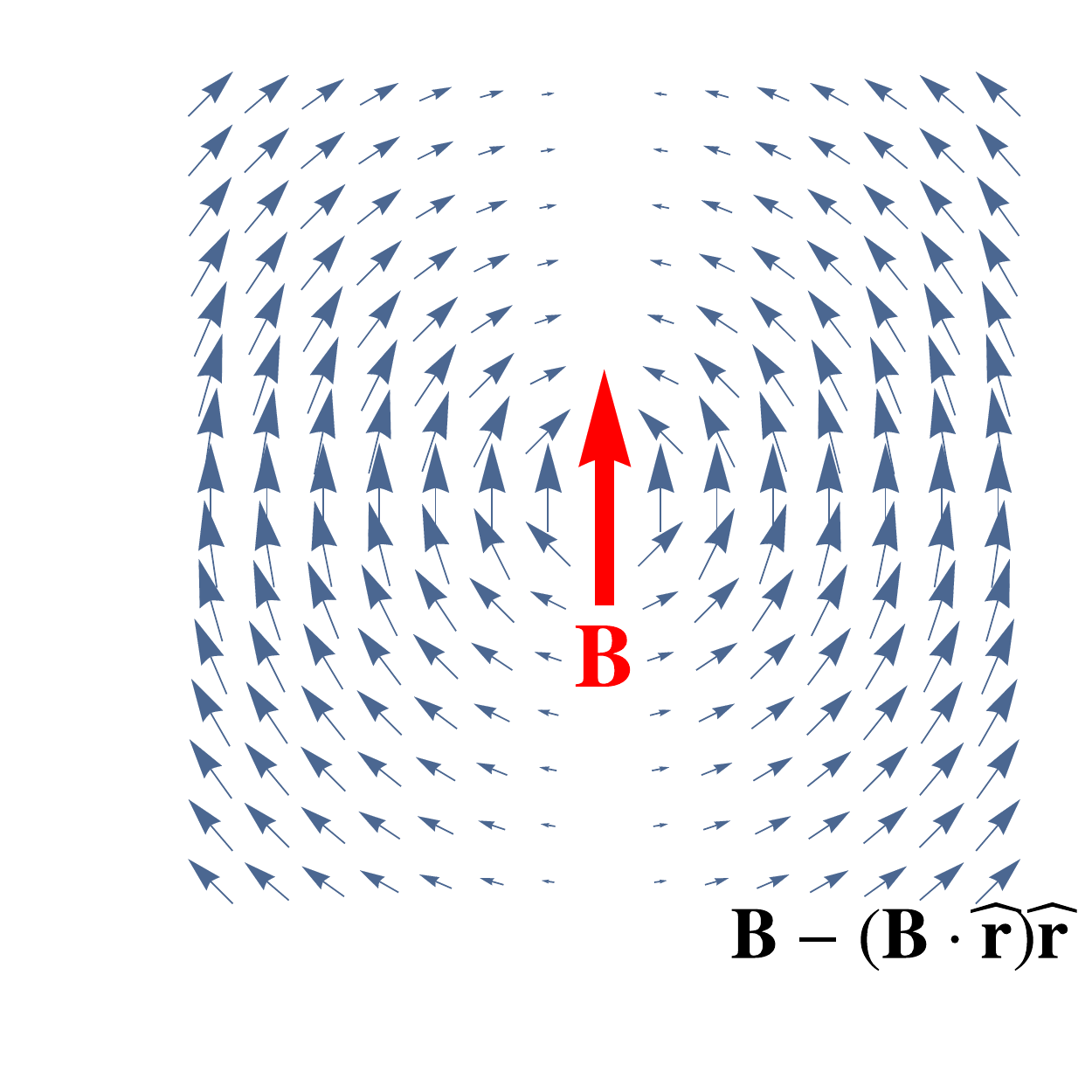}
	\end{tabular}
	\caption{The three different vectorial structures of the electric curent for a point-like source of axial imbalance. The two dimensional slices are parallel to the magnetic field $\ve{B}$ --- indicated on the right panel, pointing to the $y$-direction --- and containing the origin where the imbalance is located: $\ve{r}=(x,y)$. The three panels depict the fields $\ve{B}+(\ve{B}\cdot\widehat{\ve{r}})\widehat{\ve{r}}$, $3\left(\ve{B}\cdot\widehat{\ve{r}}\right)\widehat{\ve{r}}-\ve{B}$ and $\ve{B}-(\ve{B}\cdot\widehat{\ve{r}})\widehat{\ve{r}}$ from left to right, respectively.}\label{fig:vecFields}
\end{figure}

\section{Other side-calcs}\label{app:sidecalcs}

\subsection{Averaged longtime current of quenched local source}\label{app:avrJquenched}
Here we detail the steps leading to the final form of \mbox{Eq. (\ref{constBrespQuenchedLTAvr})}.
\begin{align}
\overline{\ve{J}} =&\actSign\frac{e^2}{2\pi^2}\ChIm\frac{1}{\pi}\frac{2}{3}\ve{B}\intlim{Q}{0}{\infty}F(q/(RT))\frac{\sin Q -Q\cos Q}{Q} = \frac{e^2}{2\pi^2}\ChIm\frac{\ve{B}}{3}\frac{2}{\pi}\underbrace{ \intlim{Q}{0}{\infty}\frac{\sin Q -Q\cos Q}{Q}\left( 1 +\intlim{\tau}{0}{\infty}\frac{\partial}{\partial \tau}\left(\frac{\sin \frac{\tau Q}{RT}}{\frac{\tau Q}{RT}}\right)f(\tau)\right). }_{=:I(RT)}
\end{align}
To evaluate the integral $I(\alpha)$, some of the steps need regularization. If needed, we insert a factor of $e^{-\epsilon Q}$ ($\epsilon >0$) or the same for $\tau$, and send the regulator $\epsilon$ to zero in the end of the calculation. In this way both integration can be done and leading us to the following simple result:

\begin{align}
I(\alpha) =& \intlim{Q}{0}{\infty}\left(-Q\frac{\partial}{\partial Q}\frac{\sin Q}{Q}\right)\left( 1+\intlim{\tau}{0}{\infty}\frac{\partial}{\partial\tau}\left(\frac{\sin \frac{\tau Q}{\alpha}}{\frac{\tau Q}{\alpha}}\right)f(\tau)\right) =\\
=& \frac{\pi}{2} +\intlim{Q}{0}{\infty}\frac{\sin Q}{Q}\intlim{\tau}{0}{\infty}\frac{\partial}{\partial\tau}\left(\cos\frac{\tau Q}{\alpha}\right)f(\tau) =\frac{\pi}{2} -\intlim{Q}{0}{\infty}\sin Q \intlim{\tau}{0}{\infty}\frac{1}{\alpha}\sin\left(\frac{\tau Q}{\alpha}\right) f(\tau) =\\
=& \frac{\pi}{2} -\frac{1}{2}\left.\intlim{x}{0}{\infty}f(x\alpha)\left(\frac{\epsilon}{(x-1)^2+\epsilon^2}-\frac{\epsilon}{(x+1)^2+\epsilon^2}\right)\right|_{\epsilon\rightarrow 0} = \frac{\pi}{2}\left( 1 -f(\alpha)\right).
\end{align}
Returning to the average-current expression:

\begin{align}
\overline{\ve{J}} =& \actSign\frac{e^2}{2\pi^2}\ChIm\ve{B}\frac{1-f(RT)}{3}.
\end{align}

\subsection{Charge transported by an axial imbalance impulse}\label{app:ChargeTranspMu5impulse}
In this section we elaborate on the expression of $\Delta Q$ which results in Fig. \ref{fig:deltaQvarR} in the main text at the end of Sec. \ref{secChargeAsymm}. We pick up from \mbox{Eq. (\ref{DeltaQfinite})} which we integrate over the infinite plane perpendicular to $\ve{B}$. This results in the following expression, containing only zero temperature contributions:
\remC{\begin{align}
\Delta Q(t_\text{obs.}) 
=& \actSign\frac{e^2B}{2\pi^2}\cdot\frac{1}{2\pi}\intlim{t}{-t_\text{obs.}/2}{t_\text{obs.}/2}\intlim{q}{-\infty}{\infty}\left(\OLChIm(t,q) +\intlim{t'}{-\infty}{0}\OLChIm(t+t',q)q\sin(qt')\right). \label{DQformula}
\end{align}}
\end{widetext}
Now we can plug in the impulselike pattern of $\OLChIm$ given in \mbox{Eq. (\ref{Mu5Impulse})}, which is normalized as
\remC{\begin{align}
\intlim{t}{-\infty}{\infty}\intlim{^3\ve{r}}{}{}\ChIm(t,r) =& \mu_0\tau L^3,
\end{align}
where $L^3$ is the effective volume of the source with the length scale \mbox{$L=\sqrt{\pi}R$}. In order to use the ansatz in \mbox{Eq. (\ref{Mu5Impulse})} we take its Fourier transform in the spatial variables to gain
\begin{align}
\OLChIm(t,q) =& \frac{\mu_0L^3}{\sqrt{\pi}}e^{-\frac{t^2}{\tau^2}-\frac{q^2R^2}{4}}.
\end{align}
Plugging the above formula into the first term of \mbox{Eq. (\ref{DQformula})} and taking the limit $t_\text{obs.}\rightarrow\infty$ we get the contribution of axial imbalance without gradients:
\begin{align}
-\frac{e^2B}{2\pi^2}\cdot\frac{1}{2\pi}\intlim{t}{-\infty}{\infty}\intlim{r}{0}{\infty}2\pi r \mu_5(t,r) =& -\mu_0\frac{e^2B}{2\pi^2}A\tau,
\end{align}
where we defined $A=L^2=\pi R^2$ as the effective surface area of the source perpendicular to the magnetic field direction. In the case of constant chiral chemical potential we would get the exact same expression despite the minus sign in the front. Now considering the full expression of \mbox{Eq. (\ref{DQformula})} we find the resulting transported charge normalized to the following dimensionless combination:
\begin{align}
\frac{\Delta Q}{C\tau}(\tau_o=t_\text{obs.}/\tau,\rho=R/\tau) =& \frac{\rho  e^{-\frac{\tobs^2}{4\left( \rho ^2+1\right)}} \text{erf}\left(\frac{\rho  \tobs}{2 \sqrt{\rho ^2+1}}\right)}{\sqrt{\rho ^2+1}},
\end{align}}
depending only on the dimensionless quantities $\textstyle\frac{t_\text{obs.}}{\tau}$ and $\textstyle\frac{R}{\tau}$, with $\textstyle C=-\mu_0\frac{e^2}{2\pi^2}BA$. Its behavior for various values of $R/\tau$ is shown in \mbox{Fig. \ref{fig:deltaQvarR}}. The small$-\tobs$ behavior is worth mentioning:
\remC{\begin{align}
\frac{\Delta Q}{C\tau}(\tau_o,\rho) \approx & \frac{1}{\sqrt{\pi}}\frac{\rho^2}{\left(\rho^2+1\right)^2}\tobs +\mathcal{O}(\tobs^2),
\end{align}}
again, getting a similar behavior to that caused by the CME relation \mbox{Eq. (\ref{CME})}: the transported charge grows with $\tobs$.

Our toy model has three scales: the observation time $t_\text{obs.}$, the spatial size $R$ and the pulse length $\tau$. As we have already seen before, in the case of large $t_\text{obs.}$ the transported charge goes to zero for any finite $R$ and $\tau$: \mbox{$\Delta Q(t_\text{obs.}\rightarrow\infty, \tau, R) = 0$}. This is the result we have already got for $\Delta Q$ defined in \mbox{Eq. (\ref{DeltaQInfZero})}.

\begin{widetext}

\begin{figure}[!h]
	\begin{tabular}{m{0.5\linewidth}m{0.5\linewidth}}
		\centering
		\includegraphics[width=0.85\linewidth]{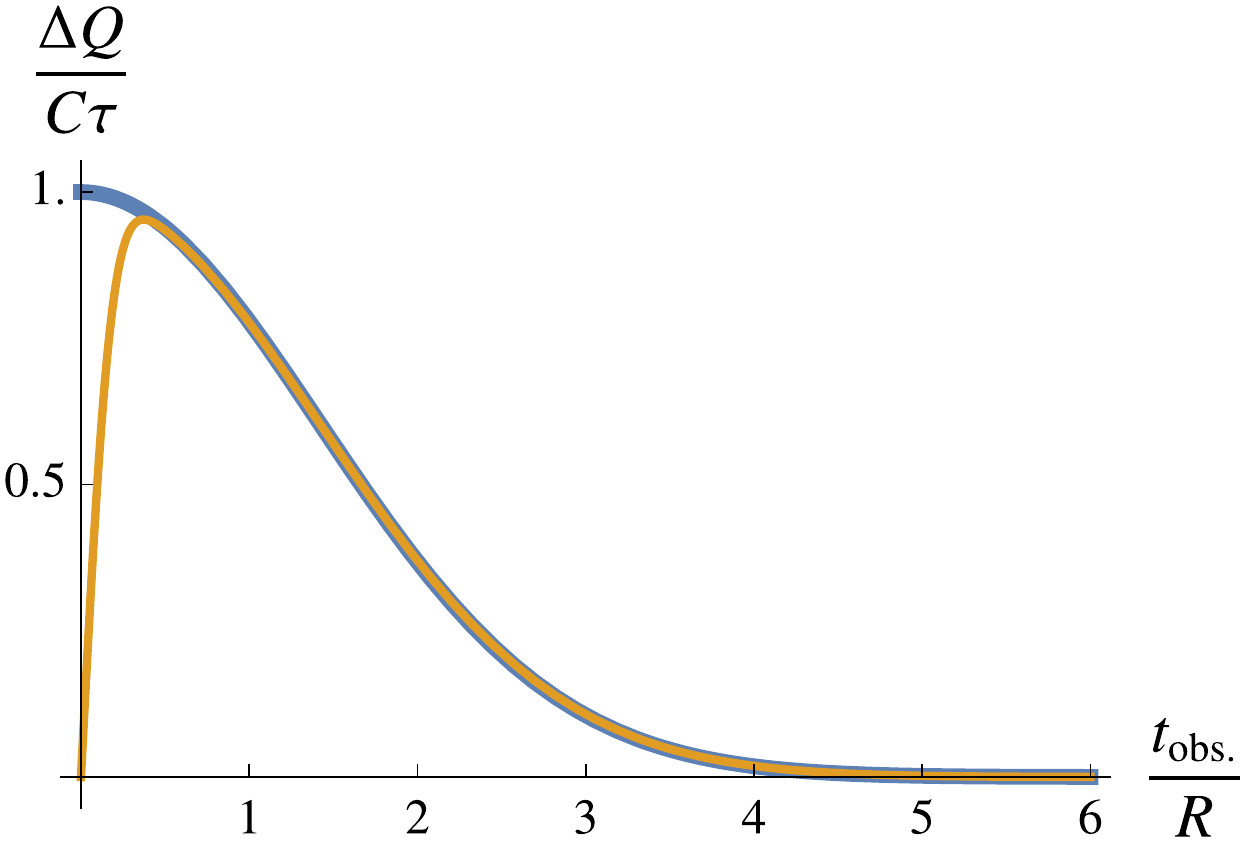} &
		\includegraphics[width=0.85\linewidth]{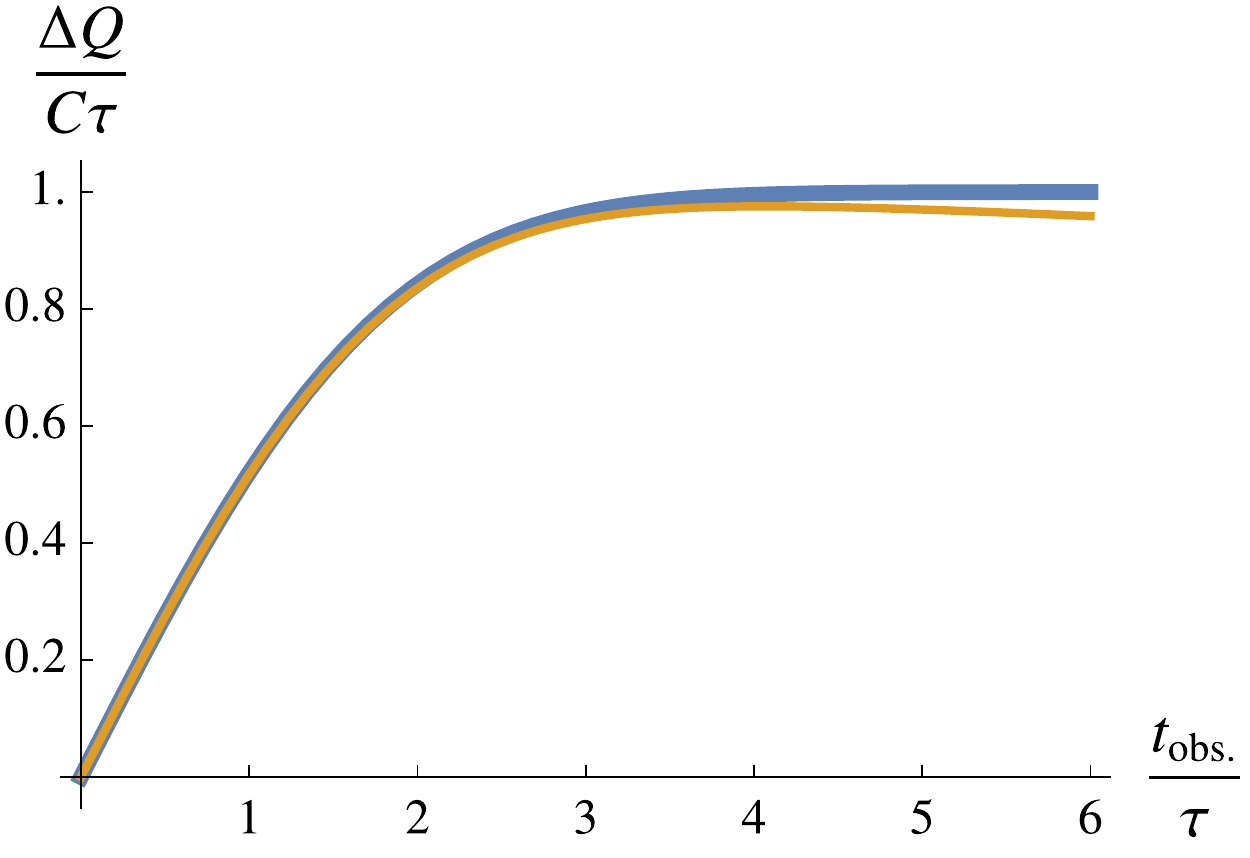} 
	\end{tabular}
	\caption{Behavior of the transported charge $\Delta Q$ over the source lifetime $\tau$. \textit{(left)} $\frac{\Delta Q}{C\tau}$ for small $\tau$, the asymptotic curve (with blue) given by \mbox{Eq. (\ref{shortImpulseMu5})}, and $\tau/R=0.1$ (with orange). \textit{(right)} $\frac{\Delta Q}{C\tau}$ in the homogeneous source limit $R\rightarrow\infty$, given by \mbox{Eq. (\ref{homMu5})}, and $R/\tau=15.0$ (with orange). For observation times larger than $\tau$ there is a saturation to the transported charge value described by the CME current in \mbox{Eq. (\ref{CME1}}.}\label{fig:transpChargeLims}
\end{figure}

\end{widetext}

For a finite length observation one can get the homogeneous source limit by sending $R$ to infinity while keeping $t_\text{obs.}$ and $\tau$ finite:
\begin{align}
\Delta Q(t_\text{obs.},\tau,R\rightarrow\infty) =& C\tau\text{erf}\left(\frac{t_\text{obs.}}{2\tau}\right), \label{homMu5}
\end{align}
as well as the short pulse limit $\tau\rightarrow 0$ with finite $t_\text{obs.}$ and $R$:
\remC{\begin{align}
\left.\frac{\Delta Q(t_\text{obs.},\tau,R)}{\tau}\right|_{\tau\rightarrow 0} =& Ce^{-\frac{t_\text{obs.}^2}{4 R^2}}. \label{shortImpulseMu5}
\end{align}}
Limiting cases are depicted in \mbox{Fig. \ref{fig:transpChargeLims}}. 

\end{document}